\documentclass[onecolumn,english,prl,showpacs,superscriptaddress,longbibliography]{revtex4-1}

\usepackage[utf8]{inputenc}
\usepackage[T1]{fontenc}
\usepackage{lmodern}
\usepackage{amsmath,amssymb,amsfonts,bm}
\usepackage{graphicx}
\usepackage{color}
\usepackage[colorlinks,citecolor=blue,linkcolor=red,urlcolor=blue]{hyperref}
\usepackage{mathrsfs}

\renewcommand{\Re}{\operatorname{Re}}

\newcommand{\dd}{\mathrm{d}}
\newcommand{\st}{\mathrm{st}}

\hypersetup{pdftitle={Black Hole Horizon Fluctuations from Unruh Quantum Noise},pdfauthor={Kashif Ammar Yasir}}
\begin{document}
	
	\title{Black Hole Horizon Fluctuations from Unruh Quantum Noise}
		\author{Kashif Ammar Yasir}
	\email{kayasir@zjnu.edu.cn}
	\affiliation{Department of Physics, Zhejiang Normal University, Jinhua 321004, China}
	\affiliation{Zhejiang Institute of Photoelectronics, Jinhua 321004, China}
	
	\date{1 October 2026}
	
\begin{abstract}
The mean Hawking flux does not determine how quantum stress fluctuations perturb a black-hole geometry. That information resides in field correlations, including vacuum contributions absent from the outgoing luminosity. Here we connect the complete exterior Unruh Wightman function to a finite-band stochastic reconstruction of Schwarzschild geometry. We retain the incoming vacuum and both outgoing frequency branches, form the connected stress covariance before angular projection, and propagate its monopole through a density-driven radial response. A shared Gaussian reconstruction then resolves how the value and derivative of the lapse control different geometric observables. For the specified covariance and noise amplitude, 2,200 realizations give relative widths of 0.0185 for the local horizon radius and 0.0483 for the surface-gravity temperature parameter, consistently across histograms and reconstruction checks. Covariance derivatives explain this unequal sensitivity. The construction separates the quantum-state input from the gravitational closure and spatial sampling assumptions, providing a reproducible finite-resolution description of how Hawking-field correlations affect horizon positions and surface-gravity statistics.
\end{abstract}
\setlength{\parskip}{0pt}
\setlength{\belowcaptionskip}{-10pt}
\maketitle

\section*{Introduction}
A black hole radiates, but its mean radiation spectrum contains only part of the information carried by the quantum field. Hawking's result relates the outgoing particle flux to the surface gravity, with curvature-dependent transmission probabilities selecting which modes reach infinity \cite{Bekenstein1973,Bardeen1973,Hawking1975,Unruh1976,Page1976,Sanchez1978}. The fluctuations of the stress tensor pose a different question: which correlations of that field are relevant to the geometry, and how do distinct geometric observables respond to them? A mean luminosity cannot answer this question, because it does not specify either the vacuum contribution to the stress noise or the correlations between different radii.

The open-system description of semiclassical gravity relates quantum stress fluctuations to a noise kernel and a causal gravitational response \cite{FeynmanVernon1963,HuMatacz1995,MartinVerdaguer1999,HuVerdaguer2008,Sinha2003,HuRoura2007}. For a radiating black hole, the central input is the state-dependent two-point function of the Hawking field. Its derivatives determine stress correlations \cite{PhillipsHu2001,Eftekharzadeh2012}, making the phase and frequency structure of the field essential even when the mean luminosity is accurately known. We use the exterior Unruh state to retain this structure explicitly and follow its consequences for geometric fluctuations.

The separation of reservoir fluctuations from system response is familiar from quantum Brownian motion and quantum measurement theory \cite{CaldeiraLeggett1983,BreuerPetruccione2002,Clerk2010}. Input--output theory and cavity optomechanics give concrete realizations of this structure \cite{GardinerCollett1985,Aspelmeyer2014}. Related open-system techniques describe spin--orbit-coupled atomic cavities \cite{Yasir2022NPJQI}, Rydberg cavity systems \cite{Yasir2026Book}, and non-Hermitian cavity magnomechanics \cite{Yu2025NPJQM,Oumie2026PRA}. Here the environment is the quantum field on the black-hole exterior: its state and stress coupling determine the source, while the gravitational response specifies how that source affects the geometry.

The Unruh state supplies the appropriate stationary exterior reference for late-time emission from a black hole formed by collapse. Its incoming sector is in vacuum, whereas its outgoing sector has spontaneous and thermally populated contributions. This state is not in global thermal equilibrium. Its Wightman function and mode-based observables have been studied numerically \cite{Hodgkinson2014}; the open problem addressed here is how to carry the complete branch structure through a specified stress projection and into a transparent reconstruction of geometric statistics. Dropping the incoming vacuum term, or replacing the outgoing sector by a Planck-weighted intensity alone, changes the stress covariance before any gravitational response is applied.

Our contribution is an explicit finite-resolution connection between these stages. We construct the complete exterior Unruh two-point function within a stated mode band, form the connected stress correlator, project the quadratic observable onto its angular monopole, and evaluate a bilocal radial response. We then use a common, explicitly specified Gaussian covariance model to compare reconstructed lapse roots with their slopes. The distributions, perturbative comparisons and sampling tests use the same realizations. This isolates the different covariance information needed by a horizon-position statistic and a surface-gravity statistic. The resulting reduced stochastic model on a Schwarzschild reference makes the state-dependent input, gravitational closure and spatial sampling explicit. It provides a calculable connection between quantum-field correlations and the distinct geometric information encoded in the lapse and its derivative.

\section*{System Modelling And Physical Idea}
We use a Schwarzschild reference geometry, \(F_0(r)=1-2GM/r\), with \(r_H=2GM\) and \(T_H=(8\pi GM)^{-1}\). Natural units \(c=\hbar=k_{\rm B}=1\) are understood; the figures also set \(G=M=1\). The field is a real, massless scalar satisfying \((\Box-\zeta R)\Phi=0\). Although \(R=0\) makes its Schwarzschild mode equation independent of the curvature coupling \(\zeta\), the stress tensor still depends on that coupling. The stress calculation uses \(\zeta=1/6\).

Let \(f^A_{\omega\ell m}=e^{-i\omega t}u^A_{\omega\ell}(r)Y_{\ell m}(\vartheta,\varphi)/(r\sqrt{4\pi\omega})\), with \(A=\mathrm{in},\mathrm{up}\), denote modes of unit incident amplitude on their respective past boundaries. The exterior Wightman function is
\begin{equation}
G_U^+(x,x')=\sum_{\ell m}\int_0^\infty\!\dd\omega\,
\left[f^{\rm in}_{\omega\ell m}(x)f^{{\rm in}*}_{\omega\ell m}(x')
 +(1+n_H)f^{\rm up}_{\omega\ell m}(x)f^{{\rm up}*}_{\omega\ell m}(x')
 +n_H f^{{\rm up}*}_{\omega\ell m}(x)f^{\rm up}_{\omega\ell m}(x')\right].
\label{eq:WU}
\end{equation}
Here \(n_H(\omega)=(e^{\omega/T_H}-1)^{-1}\). The last term has negative Killing frequency. It includes both conjugation of the spatial mode and reversal of the time dependence. Equation~\eqref{eq:WU} specifies a two-point function in the Unruh state, not a new state distinct from it \cite{Unruh1976,BirrellDavies1982,Hodgkinson2014}.

\begin{figure*}[t]
\centering
\includegraphics[width=\textwidth]{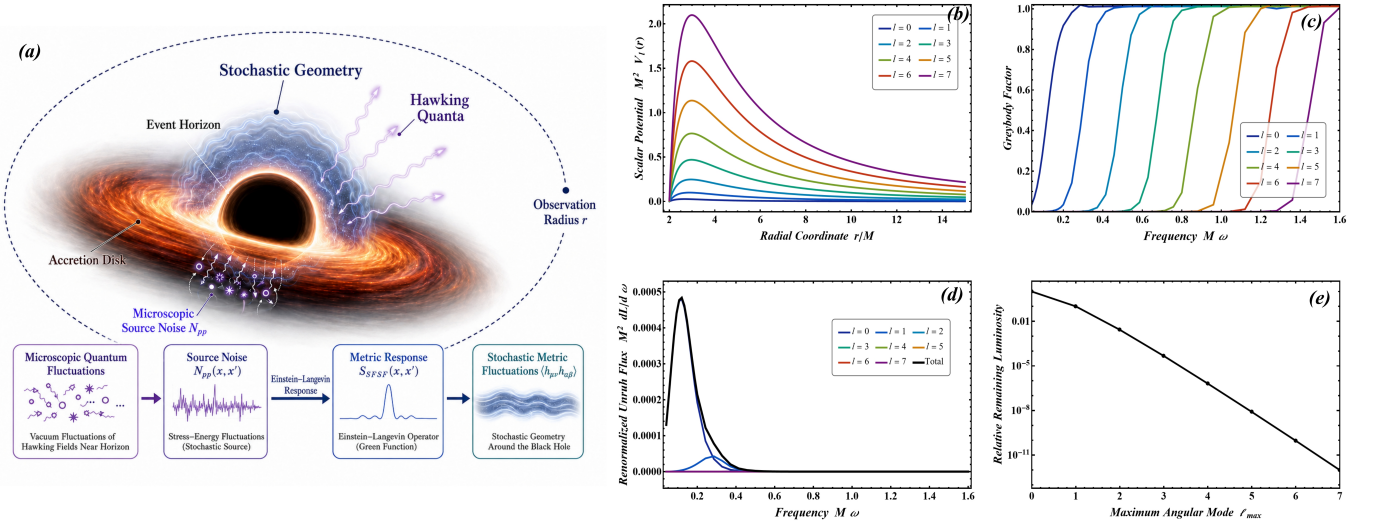}
\caption{\textbf{Field modes and the radiative channel.}
\textbf{(a)} Conceptual illustration of the field-to-geometry construction. The accretion disk is illustrative and is not included in the spherical calculation. The Einstein--Langevin arrow denotes the general framework; the numerical response studied here is its reduced density-driven model. The horizon observable below is a local lapse root, rather than a reconstructed global event horizon.
\textbf{(b)} Schwarzschild scalar potential \(M^2V_\ell(r)\) for \(\ell=0,\ldots,7\).
\textbf{(c)} Greybody transmission factors.
\textbf{(d)} Partial-wave contributions and total outgoing Unruh luminosity spectrum. The total spectrum peaks at about \(5\times10^{-4}\) in the plotted units.
\textbf{(e)} Remaining resolved luminosity fraction in the finite frequency band. Its nonzero endpoint is the last retained angular channel, not a measured contribution from uncomputed modes. This diagnostic concerns outgoing luminosity, not ultraviolet or angular convergence of the stress noise.}
\label{fig1}
\end{figure*}

The stress noise is defined by
\begin{equation}
N_{\mu\nu\alpha'\beta'}(x,x')=
\frac12\left\langle\{\hat t_{\mu\nu}(x),\hat t_{\alpha'\beta'}(x')\}\right\rangle_U,
\qquad \hat t_{\mu\nu}=\hat T_{\mu\nu}-\langle\hat T_{\mu\nu}\rangle_U .
\label{eq:noise}
\end{equation}
For the free field, Wick's theorem expresses this quantity as derivatives of products of \(G_U^+\). Subtracting the mean stress removes the disconnected term; it does not remove vacuum stress fluctuations. The finite-band calculation therefore represents the full Unruh covariance within its retained band. It does not calculate an excess covariance relative to the Boulware state, and a band cutoff is not a substitute for covariant renormalization of a coincidence limit.

In the full theory, the metric perturbation obeys an Einstein--Langevin equation of the form
\begin{equation}
\mathcal D h+H_{\rm ret}*h=8\pi G\,\xi,
\qquad \langle\xi(x)\xi(x')\rangle_{\st}=N(x,x').
\label{eq:EL_main}
\end{equation}
The displayed response is more restricted. Define \(\rho=T_{\mu\nu}u^\mu u^\nu\) for radial free fall from rest at infinity and \(\rho_0=(4\pi)^{-1}\int\dd\Omega_{\boldsymbol n}\rho\). We denote its stochastic fluctuation by \(\eta\). The reduced model applies
\begin{equation}
\delta m(r)=-4\pi\int_r^{R_B}\dd s\,s^2\eta(s),
\qquad
\delta F(r)=\frac{8\pi G}{r}\int_r^{R_B}\dd s\,s^2\eta(s).
\label{eq:source_main}
\end{equation}
Thus \(\delta m(R_B)=0\). This fixes the mass at a finite numerical boundary; identifying it with a fixed ADM mass additionally requires an asymptotic completion.
For the kernel \(K(r,s)=8\pi Gs^2\Theta(s-r)\Theta(R_B-s)/r\), the bilocal response has the explicit covariance structure
\begin{equation}
S_F(\Omega;r,r')=\int\dd s\,\dd s'\,K(r,s)\widetilde N_0(\Omega;s,s')K(r',s'),\qquad
C_F^{\rm rad}=\int_0^{\Omega_{\max}}\frac{\dd\Omega}{\pi}\Re S_F.
\label{eq:cov_main}
\end{equation}
Here \(\widetilde N_0\) is the spectrum of the stress monopole. The double radial integral is needed even for \(S_F(\Omega;r,r)\); a local source spectrum alone does not determine a metric variance.

The choice of density in Eq.~\eqref{eq:source_main} is a closure assumption. In ingoing Eddington--Finkelstein coordinates,
\begin{equation}
\dd s^2=-e^{2\psi}F\,\dd v^2+2e^\psi\dd v\dd r+r^2\dd\Omega^2,
\qquad F=1-2Gm/r,
\label{eq:metric_main}
\end{equation}
the Einstein equation gives \(\partial_r m=-4\pi r^2T^v{}_v\). A freely falling density is generally different from \(-T^v{}_v\). Its insertion into Eq.~\eqref{eq:source_main} does not by itself solve the remaining flux and redshift equations. We therefore refer to the displayed quantity as a density-driven constraint response. The retarded polarization kernel in Eq.~\eqref{eq:EL_main} has not been included in this response.

The geometry reconstruction sets \(\psi=0\) and uses \(F_\xi=\bar F+\delta F_\xi\). A simple root defines a candidate trapping horizon \cite{Hayward1994,AshtekarKrishnan2004,NielsenVisser2006}, and the associated slope defines a surface-gravity temperature parameter,
\begin{equation}
F_\xi(r_h^{(\xi)})=0,
\qquad T_\xi=\frac{F_\xi'(r_h^{(\xi)})}{4\pi}.
\label{eq:horizon_temp_main}
\end{equation}
For a slowly evolving horizon this has a geometric thermodynamic interpretation \cite{Kodama1980,Hayward1998}. The propagation of quantum fields through fluctuating geometries introduces further correlations \cite{Barrabes2000}. A fluctuating slope alone does not determine the particle spectrum seen at infinity. A redshift sector, time evolution and a field-propagation calculation are needed to make that additional identification.

\section*{Results}
\subsection*{Microscopic field input and what its diagnostics establish}
The potential in Fig.~\ref{fig1}(b) raises the transmission threshold as angular momentum increases. The greybody factors in Fig.~\ref{fig1}(c) show this progression, and the Hawking occupation suppresses the high-frequency contribution to the mean luminosity in Fig.~\ref{fig1}(d). Low angular momenta dominate the outgoing flux. The tail in Fig.~\ref{fig1}(e) consequently falls rapidly over the computed channels.

These radiative observations do not certify the stress kernel. The \(\mathrm{in}\) vacuum and the spontaneous part of the \(\mathrm{up}\) sector in Eq.~\eqref{eq:WU} are not multiplied by a decaying Planck occupation. Higher frequencies and angular momenta can therefore matter much more for local stress correlations than for the luminosity. Supplementary Fig.~S1 records the numerical accuracy of the current scattering basis: the flux residual reaches order \(10^{-2}\), while the fit residual is typically several \(10^{-3}\), with larger isolated excursions. These data support a qualitative mode-resolved analysis, but do not establish a precision bound on the derivative-dependent stress noise.

The coherence curves in Supplementary Fig.~S1 retain all three Unruh terms at equal time for each positive-frequency shell. The nearest radial pair has the largest low-frequency field coherence. Both field and gradient coherences exhibit minima and revivals as frequency varies; neither is a monotone measure of spatial separation. Their bounds follow from positivity of the retained two-point function, whereas the locations and depths of their minima remain subject to matching and grid errors.

\subsection*{Finite-band source and density-driven response}
\begin{figure*}[t]
\centering
\includegraphics[width=\textwidth]{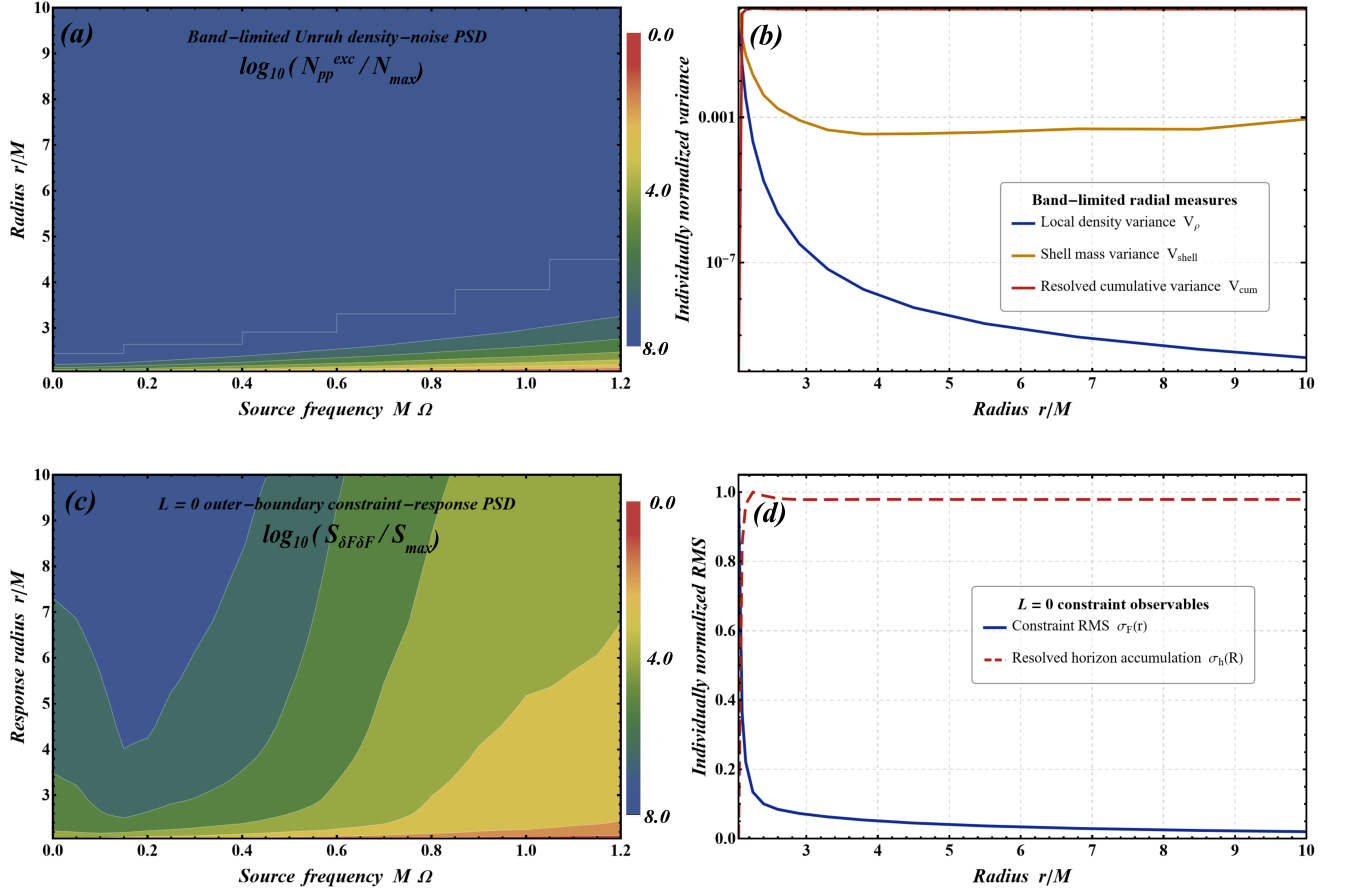}
\caption{\textbf{Band-limited Unruh density noise and its reduced spherical response.}
\textbf{(a)} Local density-noise PSD for \(0\le M\Omega\le1.2\) and \(2.05\le r/M\le10\). The quantity is the full retained-band Unruh noise; the superscript ``exc'' in the supplied artwork is a legacy label and does not denote a vacuum subtraction.
\textbf{(b)} Local density variance, cell-weighted monopole shell mass variance and cumulative variance, normalized separately.
\textbf{(c)} PSD obtained from the \(L=0\) density-driven radial map with \(\delta m(30M)=0\). This is a constraint-model spectrum, not a computed retarded polarization response.
\textbf{(d)} Separately normalized lapse RMS and cumulative near-horizon diagnostic. Both use the resolved radial domain. In \textbf{a,c}, colours encode the relative logarithm from \(-8\) to \(0\); the printed positive numbers 4 and 8 denote the magnitudes of negative logarithmic levels. The finite inner cutoff and separate normalizations must be retained when comparing panels.}
\label{fig2}
\end{figure*}

Figure~\ref{fig2}(a) shows a strongly concentrated near-horizon source over the resolved domain. Most of the plotted region lies many decades below its maximum. This statement concerns the computed band and the inner cutoff, \(r_{\min}=2.05M\); it is not a statement about the value of a renormalized local noise kernel exactly on the horizon. The source frequency \(\Omega\) is a frequency of the quadratic stress observable, and is distinct from the field frequency \(\omega\). Positive--negative frequency products contribute at small \(\Omega\), including \(\Omega=0\).

The radial profiles in Fig.~\ref{fig2}(b) distinguish a local density variance, a shell mass variance and a resolved cumulative variance. Each curve is divided by its own maximum. The blue curve decreases rapidly, the shell measure reaches a shallow minimum and then changes more slowly, and the cumulative measure rapidly approaches a plateau. The shell measure includes the radial quadrature cell width; it cannot be compared directly with the local density variance as though the two had the same dimensions or normalization.

Applying the bilocal radial map reshapes the source spectrum, as seen in Fig.~\ref{fig2}(c). Its low-frequency trough and broader frequency structure need not mirror the local source map because the response contains correlations between different radii. In Fig.~\ref{fig2}(d), the normalized lapse RMS falls away from the inner edge, while the accumulated diagnostic reaches nearly its full displayed value close to that edge. The small subsequent decrease is permissible: contributions from additional shells include cross-covariances and need not add monotonically. A nearly saturated cumulative curve is evidence about the resolved integral, not evidence that the omitted interval between \(2M\) and \(2.05M\) is negligible.

\subsection*{A common ensemble for horizon and surface-gravity statistics}
Figure~\ref{fig3}(a) shows a reconstructed lapse ensemble about the Schwarzschild reference. The pointwise uncertainty is most visible close to the zero of the lapse; farther out, the mean profile approaches its asymptotic form and the displayed fluctuations narrow. These shaded bands describe the dispersion of the lapse at fixed radius, not simultaneous confidence bands for an entire geometry. Nor is the ensemble mean lapse, which is fixed by centering the perturbations, a separately calculated mean backreaction solution.

The histograms contain 2,200 shared realizations. The normalized radius in Fig.~\ref{fig3}(b) has mean \(0.9996\) and standard deviation \(0.0185\), while the surface-gravity temperature parameter in Fig.~\ref{fig3}(c) has mean \(1.0021\) and standard deviation \(0.0483\). Thus the relative temperature width is about 2.61 times the relative radius width. These values belong to the stated smoothed covariance model at imposed amplitude \(\lambda_Q=1\). They are not absolute fluctuation amplitudes fixed solely by the Unruh state.

The origin of the unequal widths can be stated without reference to a histogram. At a simple reference root, let \(a=\bar F'(\bar r_h)\), \(b=\bar F''(\bar r_h)\), \(X=\delta F(\bar r_h)\) and \(Y=\delta F'(\bar r_h)\). To first order,
\begin{equation}
\delta r_h=-\frac{X}{a},\qquad
\delta T=\frac{Y-(b/a)X}{4\pi}.
\label{eq:linear}
\end{equation}
Writing \(C_{00}=\langle X^2\rangle\), \(C_{01}=\langle XY\rangle\) and \(C_{11}=\langle Y^2\rangle\) gives
\begin{equation}
\operatorname{Var}(\delta r_h)=\frac{C_{00}}{a^2},\qquad
\operatorname{Var}(\delta T)=\frac{C_{11}-2(b/a)C_{01}+(b/a)^2C_{00}}{16\pi^2}.
\label{eq:widths}
\end{equation}
For Schwarzschild, \(\delta r_h/r_H=-X\) and \(\delta T/T_H=2X+r_HY\). A fixed metric variance therefore does not fix the temperature variance: the value--slope covariance and the slope variance also enter. This dependence makes a temperature statistic a particularly sensitive probe of radial smoothing and derivative resolution.

The geometry ensemble uses a Gaussian radial covariance whose RMS envelope is obtained from the finite-band response and then smoothed. Its amplitude and correlation length are specified in Methods. Even before the local polynomial fit, a covariance of the form \(C_F(r,r')=s(r)s(r')\exp[-(r-r')^2/(2L_c^2)]\) yields
\begin{equation}
\operatorname{Var}(\delta T)=\frac{[s'-(b/a)s]^2+s^2/L_c^2}{16\pi^2}
\quad\text{at }r=\bar r_h.
\label{eq:envelope_sensitivity}
\end{equation}
Equation~\eqref{eq:envelope_sensitivity} displays separately the envelope gradient and the correlation-length contribution. The implemented quadratic fit replaces these point derivatives by finite-window linear functionals, whose covariance is evaluated explicitly. Matching only the noise amplitude at one radial point would not ensure matching temperature distributions.

Both widths grow nearly linearly with \(\lambda_Q\) in Fig.~\ref{fig3}(d). This agrees with the leading-order consequence of \(C_F\propto\lambda_Q^2\). The same underlying random coefficients are used throughout the amplitude scan, so changes between amplitudes are not obscured by independently redrawing the noise. The small offsets of the histogram means from unity are compatible with the nonlinear root and slope maps: a centered lapse perturbation does not imply centered nonlinear observables. Their signs and magnitudes are model-dependent and are not identified with a renormalized mean quantum correction.

\subsection*{Paired reconstruction checks}
Extended Data Fig.~\ref{edfig:metric_validation} examines the same ensemble. Its covariance panel displays the Gaussian reconstruction model, not the unrestricted microscopic Unruh stress kernel. The highest displayed covariance occurs near the smaller radii in the shown window and decreases toward larger radii; its appearance reflects both the decaying RMS envelope and the imposed correlation kernel. It cannot be used alone to extract a correlation length from the width of a diagonal feature.

In Extended Data Fig.~\ref{edfig:metric_validation}(b), the linearized and root-evaluated relative radius widths are \(0.0184\) and \(0.0185\). The corresponding temperature widths in panel~c are \(0.0482\) and \(0.0483\). The near-diagonal point clouds show that the linear approximation captures the principal variation across this ensemble, while the systematic curvature and residual spread expose the nonlinear corrections. The quoted changes are about 0.5\% and 0.2\%, respectively, at the precision printed in the panels. These rounded width comparisons are not error bounds on individual realizations. The root-evaluated values agree with the histograms because the same arrays define both figures; this is a consistency check, not an independent replication.

Panel~d follows estimates from nested prefixes of the 2,200-sample ensemble. The temperature estimate changes appreciably at the smallest sample counts and then settles near the reported value; the radius estimate is less variable on the same vertical scale. For independent approximately Gaussian observables, the relative sampling uncertainty of a standard deviation is about \([2(N_s-1)]^{-1/2}\), or 1.5\% at \(N_s=2200\). This supplies an indicative statistical scale, not a bootstrap confidence interval for the weakly non-Gaussian transformed observables. The close linear/root comparison benefits from pairing the realizations. Neither pairing nor sampling stability tests the field-frequency band, angular truncation, radial matching or the chosen continuation through the reference horizon.

\section*{Discussion}
\begin{figure*}[t]
\centering
\includegraphics[width=\textwidth]{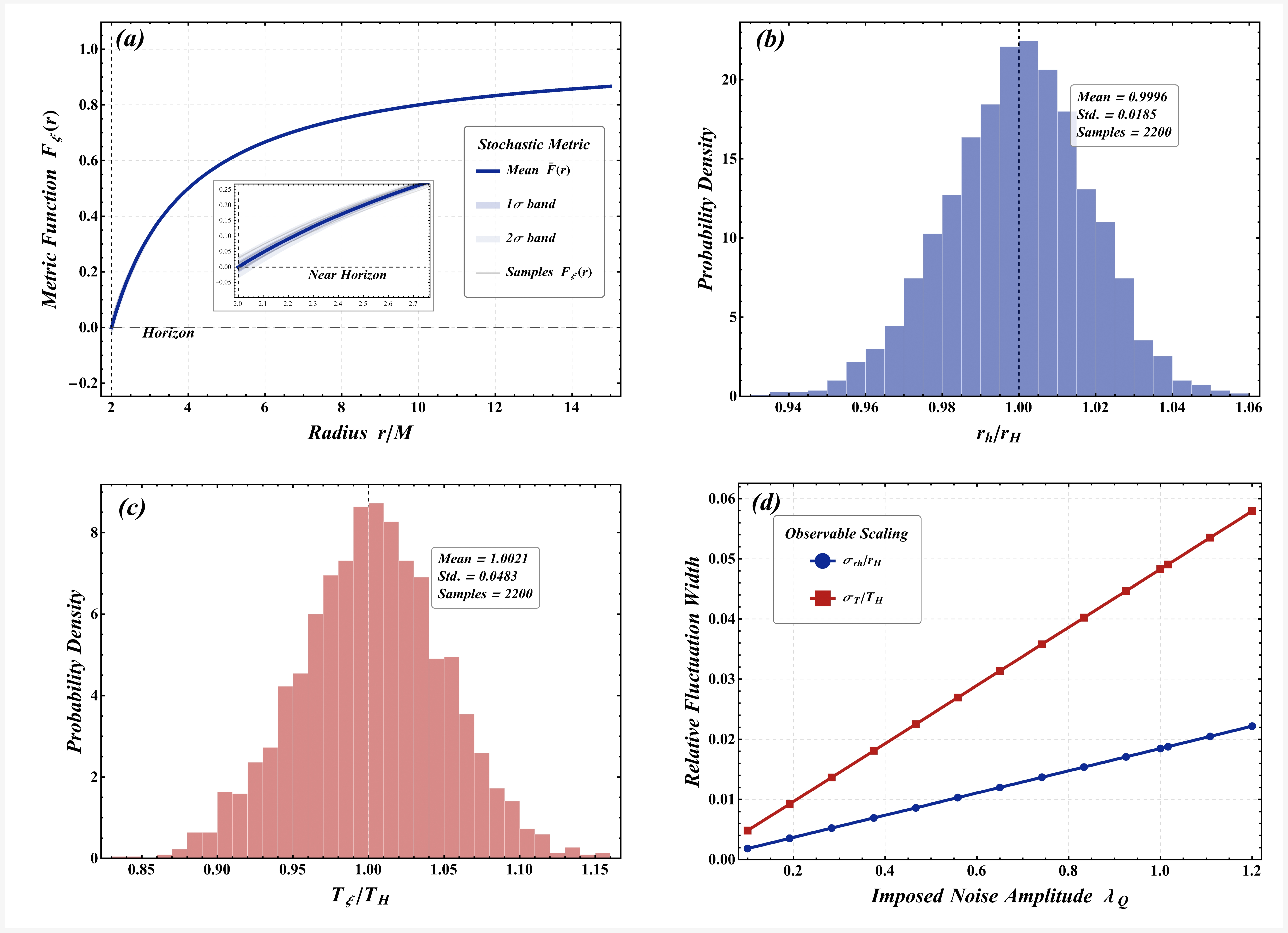}
\caption{\textbf{Shared stochastic lapse ensemble and geometric distributions.}
\textbf{(a)} Mean lapse, pointwise one- and two-standard-deviation bands and representative realizations; the inset resolves the reference-horizon region.
\textbf{(b)} Root-evaluated radius distribution for 2,200 realizations: mean \(0.9996\), relative standard deviation \(0.0185\).
\textbf{(c)} Surface-gravity temperature parameter from the same local polynomial realization evaluated at its root: mean \(1.0021\), relative standard deviation \(0.0483\). This is a geometric slope statistic, not a computed radiation-temperature distribution at infinity.
\textbf{(d)} Relative widths versus the imposed amplitude \(\lambda_Q\), using paired realizations throughout the scan. The histogram values correspond to \(\lambda_Q=1\). The Gaussian correlation length and RMS-envelope smoothing length are both \(0.75M\); the unprojected RMS envelope is normalized to \(0.018\) at the reference horizon. The finite-window fit and the finite ensemble do not force the measured radius width to equal this input amplitude.}
\label{fig3}
\end{figure*}
A stochastic description of black-hole geometry requires more information than the Hawking luminosity. The mean outgoing flux is thermally weighted, whereas stress fluctuations also contain vacuum terms and cross-products of the Wightman branches. Retaining these terms before the stress monopole is formed preserves correlations that are lost in a thermal-intensity model. The radial response then combines correlations between source shells, so a concentrated local spectrum need not give a response with the same radial or frequency dependence.

The shared reconstruction adds a second distinction. Horizon displacement probes a metric value, whereas surface gravity probes a correlated value and derivative. Equations~\eqref{eq:widths} and \eqref{eq:envelope_sensitivity} identify the quantities that control their relative spreads. For the displayed reconstruction, the temperature width is larger, but this ordering is not universal. Changing the correlation length or smoothing prescription can alter the derivative covariance without proportionally changing the radius variance. Reporting both observables together therefore provides a more informative test of a stochastic geometry than reporting either width alone.

The role of the Gaussian reconstruction must remain explicit. The microscopic density noise is derived from the finite-band Unruh modes, but the spatial covariance used to draw the final lapse ensemble is a model built from a smoothed diagonal RMS profile and an imposed Gaussian correlation kernel. Its positivity is constructed; that property does not independently establish positivity or convergence of the numerical microscopic unequal-radius kernel. Likewise, Gaussian lapse sampling reproduces the selected covariance but not the higher cumulants of the quadratic quantum stress. The agreement of Fig.~\ref{fig3} and Extended Data Fig.~\ref{edfig:metric_validation} now establishes a consistent comparison within this model.

The extension to dynamical backreaction is well defined but not completed here. It requires the tensor source components entering the gravitational constraints, a compatible redshift and flux sector, and the causal stress-polarization response. It also requires a spacetime sampling prescription and control of the near-horizon and mode-cutoff limits \cite{Sinha2003,HuRoura2007,HuVerdaguer2008}. The scattering diagnostics supplied with the field figure remain at the percent level in parts of the grid. The separate higher-order source solver used for the shared reconstruction has its own diagnostics; the updated artwork does not report them and cannot certify their convergence. These issues set the accuracy and physical scope of the numerical widths, rather than being removed by cross-figure agreement.

\section*{Conclusion}
We have connected the complete exterior Unruh Wightman structure to finite-band stress correlations, a reduced spherical response and a common stochastic reconstruction of black-hole geometry. The calculation distinguishes quantum-state information from the assumptions used to convert it into spatial metric samples. In the specified ensemble, the root and surface-gravity widths are \(0.0185\) and \(0.0483\), and paired linear/root comparisons resolve small nonlinear corrections. The principal physical result is the different covariance sensitivity of horizon position and surface gravity, with radial derivatives controlling the latter. This formulation provides a concrete basis for testing how state choice, spatial resolution and gravitational response affect geometric fluctuations. A prediction of an absolute or regulator-independent black-hole width will require the conserved tensor response and a converged sampling prescription in addition to the construction presented here.

\section*{Acknowledgments}
K.A.Y. acknowledges support from the Research Fund for International Young Scientists of the NSFC under Grant No. KYZ04Y22050, Zhejiang Normal University research funding under grant No. ZC304021914 and Zhejiang province postdoctoral research project under grant number ZC304021952.

\section*{Author contributions}
K.A.Y. conceived the project, developed the theoretical framework, performed the calculations, generated the figures and wrote the manuscript.

\section*{Competing interests}
The author declares no competing interests.
	
%%%%%%%%%%%%%%%%%%%%%%%%%%%%%%%%%%%%%%%%%%%%%%%%%%%%%%%%%%%%%%%%%%%%%
\section*{Methods}
\subsection*{Radial modes and boundary normalization}
With \(\dd r_*/\dd r=F_0^{-1}\), the scalar radial equation is
\begin{equation}
\left[\frac{\dd^2}{\dd r_*^2}+\omega^2-V_\ell(r)\right]u_{\omega\ell}=0,
\qquad V_\ell=F_0\left[\frac{\ell(\ell+1)}{r^2}+\frac{2GM}{r^3}\right].
\end{equation}
The horizon-ingoing solution \(h\) is normalized by \(h\sim e^{-i\omega r_*}\) at the future horizon and matched to \(A_{\rm in}e^{-i\omega r_*}+A_{\rm out}e^{i\omega r_*}\) at infinity. The modes entering Eq.~\eqref{eq:WU} are
\begin{equation}
u^{\rm in}=\frac{h}{A_{\rm in}},\qquad
u^{\rm up}=h^*-\frac{A_{\rm out}^*}{A_{\rm in}}h.
\label{eq:normalization}
\end{equation}
The second expression has unit incident amplitude from the past horizon. Its transmitted amplitude at infinity is \(1/A_{\rm in}\), using the exact Wronskian identity \(|A_{\rm in}|^2-|A_{\rm out}|^2=1\). Numerical errors in that identity must be measured before using it as an accuracy claim. In particular, the independent quantities
\begin{equation}
\Gamma_{\omega\ell}=|A_{\rm in}|^{-2},\qquad
\mathcal R_{\omega\ell}=|A_{\rm out}/A_{\rm in}|^2,\qquad
\epsilon_{\rm flux}=|\Gamma_{\omega\ell}+\mathcal R_{\omega\ell}-1|
\end{equation}
are used rather than defining \(\Gamma=1-\mathcal R\) and then testing their sum.

The mean outgoing energy spectrum at infinity is
\begin{equation}
\frac{\dd L}{\dd\omega}=\sum_\ell\frac{2\ell+1}{2\pi}\,
\omega\Gamma_{\omega\ell}n_H(\omega).
\end{equation}
It contains no separate negative-frequency outgoing energy flux. The negative-frequency term of Eq.~\eqref{eq:WU} is instead necessary for field and stress correlation functions. In a finite angular calculation, a resolved luminosity remainder can be formed by summing the retained channels from a selected index to \(\ell=7\). A nonzero value at the final index represents that final retained channel; it is not an estimate of the uncomputed \(\ell>7\) tail.

\subsection*{Stress observable, Wick contraction and signed frequencies}
The freely falling frame uses \(u^t=F_0^{-1}\), \(u^r=-\sqrt{2GM/r}\). On the Ricci-flat background, the on-shell scalar density in this geodesic frame can be written
\begin{equation}
\rho=\frac12(D_u\Phi)^2+\left(\frac12-2\zeta\right)\sum_{i=1}^3(D_i\Phi)^2
       -2\zeta\Phi D_u^2\Phi,\qquad \zeta=\frac16,
\label{eq:rho_methods}
\end{equation}
where products are symmetrized in the point-split quantum operator. The spatial derivatives refer to an orthonormal frame. The geodesic condition is needed to identify \(D_u^2\Phi\) with \(u^\mu u^\nu\nabla_\mu\nabla_\nu\Phi\).

Let \(X=(\Phi,D_u\Phi,D_1\Phi,D_2\Phi,D_3\Phi,D_u^2\Phi)\) and write \(\rho=C_{ab}X_aX_b\), with symmetric \(C\). Its nonzero entries are \(C_{22}=1/2\), \(C_{33}=C_{44}=C_{55}=1/2-2\zeta\), and \(C_{16}=C_{61}=-\zeta\). If \(W_{ac}(x,x')=\langle X_a(x)X_c(x')\rangle_U\), the connected density correlator is
\begin{equation}
C_\rho^>(x,x')=C_{ab}C_{cd}\big[W_{ac}W_{bd}+W_{ad}W_{bc}\big].
\label{eq:wick_methods}
\end{equation}
This formula retains the cross-products between all field sectors. Using only a thermally weighted up mode, or applying Wick's theorem to a vacuum-subtracted Wightman function, does not give the full Unruh noise \cite{PhillipsHu2001}.

Our Fourier convention is \(W(\tau)=\int\dd\nu\,e^{-i\nu\tau}\mathcal W(\nu)\) and \(\widetilde C(\Omega)=\int\dd\tau\,e^{i\Omega\tau}C(\tau)\). Equation~\eqref{eq:wick_methods} becomes a convolution of \(\mathcal W(\nu)\) and \(\mathcal W(\Omega-\nu)\), with an overall factor \(2\pi\). Both signed-frequency supports must be included. The symmetrized bilocal spectrum is
\begin{equation}
\widetilde N_{ij}(\Omega)=\frac12\left[\widetilde C^>_{ij}(\Omega)+\widetilde C^>_{ji}(-\Omega)\right].
\end{equation}
Exchanging the spatial arguments is essential; a bilocal spectrum need not be real. For an autocorrelation, the band-limited variance is \(\int_0^{\Omega_{\max}}\dd\Omega\,\widetilde N_{ii}(\Omega)/\pi\). The same integral with the real part gives the equal-time, symmetrized cross-covariance. The lower endpoint is retained because opposite-sign field frequencies can give zero stress frequency.

\subsection*{Angular projection and radial covariance}
The local spectrum evaluates both angular arguments at coincidence. The monopole used in the radial map instead averages the stress covariance over two spheres. Rotational invariance gives
\begin{equation}
\widetilde N_0(\Omega;r,r')=
\frac12\int_{-1}^{1}\dd\mu\,\widetilde N_\rho(\Omega;r,r',\mu),
\qquad \mu=\boldsymbol n\cdot\boldsymbol n'.
\end{equation}
This projection follows Eq.~\eqref{eq:wick_methods}. Keeping only the scalar-field mode \(\ell=0\) before forming the stress tensor would describe a different approximation.

For \(K(r,s)=8\pi Gs^2\Theta(s-r)\Theta(R_B-s)/r\), the response spectrum is
\begin{equation}
S_F(\Omega;r,r')=\int\dd s\,\dd s'\,K(r,s)\widetilde N_0(\Omega;s,s')K(r',s').
\label{eq:response_methods}
\end{equation}
The diagonal supplies Fig.~\ref{fig2}(c); its frequency integral supplies \(\sigma_F^2(r)\). Off-diagonal radial terms are retained. A local-noise approximation proportional to \(\delta(s-s')\) would not reproduce this calculation. Equation~\eqref{eq:response_methods} is the covariance of the specified radial closure, without the causal polarization kernel of Eq.~\eqref{eq:EL_main}.

On a radial quadrature, \(K\) includes each cell weight once. The source covariance matrix itself is an unweighted covariance at the nodes. The shell variance in Fig.~\ref{fig2}(b) is \((4\pi r_i^2w_i)^2C_{0,ii}\). The cumulative diagnostic uses weights \(8\pi Gr_i^2w_i/r_H\) over \([r_{\min},R]\). It is a variance of an integrated source, including cross-covariances; it is not an integral of the pointwise RMS. Normalized profiles divide each observable by its own maximum over the displayed domain. Logarithmic floors affect the display and do not restore missing small-scale physics.

\subsection*{Influence functional and the stochastic representation}
The underlying field theory is specified by
\begin{equation}
S_\Phi=-\frac12\int\dd^4x\sqrt{-g}\,[g^{\mu\nu}\nabla_\mu\Phi\nabla_\nu\Phi+\zeta R\Phi^2].
\end{equation}
The closed-time-path construction keeps response and fluctuations in a common nonequilibrium description \cite{CalzettaHu2008}; stochastic detector models provide a complementary field-based example \cite{RavalHuAnglin1996}. Integrating the field along the forward and backward histories defines the influence functional \(\exp(iS_{\rm IF}[g^+,g^-])=\operatorname{Tr}(U[g^+]\rho_U U[g^-]^\dagger)\). With the linear stress source denoted by \(J_{\mu\nu}=h_{\mu\nu}/2\), let \(J_\Delta=J^+-J^-\) and \(J_c=(J^++J^-)/2\). The quadratic expansion has the form
\begin{equation}
S_{\rm IF}^{(2)}=\int J_\Delta\langle T\rangle_U
 +\int J_\Delta\chi_{\rm ret}J_c
 +\frac{i}{2}\int J_\Delta N J_\Delta,
\label{eq:IF_methods}
\end{equation}
where tensor contractions and invariant measures are understood. Local metric-variation terms are included in the response kernel. With this source convention its nonlocal part is the causal commutator response. The Gaussian identity
\begin{equation}
\exp[-\tfrac12 J_\Delta N J_\Delta]
 =\left\langle\exp[i\int J_\Delta\xi]\right\rangle_{\st},
\qquad \langle\xi\rangle_{\st}=0,\quad \langle\xi\xi\rangle_{\st}=N,
\end{equation}
leads to the linearized stochastic equation
\begin{equation}
\delta\mathcal E_{\mu\nu}[h]-8\pi G\,\delta\langle T_{\mu\nu}\rangle_{\rm ret}[h]
 =8\pi G\,\xi_{\mu\nu}.
\end{equation}
Here \(\mathcal E\) includes the local gravitational terms and any required renormalized curvature terms. This is the standard Einstein--Langevin construction \cite{HuMatacz1995,MartinVerdaguer1999,HuVerdaguer2008}; our radial closure does not evaluate its full retarded inverse. A classical Gaussian source represents symmetrized second moments at this order. It is not an operator identity replacing the complete quantum stress, and global equilibrium fluctuation--dissipation relations cannot be imposed on the full Unruh state.

\subsection*{Shared covariance model for the geometric ensemble}
To distinguish the two stages, denote the covariance from Eq.~\eqref{eq:response_methods} by \(C_F^{\rm rad}\) and the Gaussian sampling model by \(C_F^{\rm mod}\). Let \(s_i^{\rm raw}=\sqrt{C_F^{\rm rad}(r_i,r_i)}\). The profile is linearly interpolated to a uniform grid and extended to \(r_H\) with the first resolved value. On that grid, define
\begin{equation}
A_{ij}=\frac{w_j\exp[-(r_i-r_j)^2/(2L_s^2)]}
 {\sum_k w_k\exp[-(r_i-r_k)^2/(2L_s^2)]},\quad
\bar s_i=\sum_jA_{ij}s_j^{\rm raw},\quad
s_i=\epsilon_F\frac{\bar s_i}{\bar s_1},
\label{eq:smoothing}
\end{equation}
where \(w_j\) are trapezoidal radial weights. The model covariance is
\begin{equation}
C_{F,ij}^{\rm mod}=\lambda_Q^2s_i s_j
\exp[-(r_i-r_j)^2/(2L_c^2)].
\label{eq:modelcov}
\end{equation}
We use \(L_s=L_c=0.75M\), \(\epsilon_F=0.018\) and \(\lambda_Q=1\) for the histograms. Smoothing an RMS envelope in Eq.~\eqref{eq:smoothing} is not the same operation as applying a linear smearing operator to the complete covariance. In particular, Eq.~\eqref{eq:modelcov} does not retain the microscopic off-diagonal entries of \(C_F^{\rm rad}\). The imposed amplitude absorbs its absolute normalization.

For \(C_F^{\rm mod}=Q\Lambda Q^{\mathsf T}\), samples are drawn as \(\delta\boldsymbol F=Q\Lambda^{1/2}\boldsymbol z\) with independent standard normal \(z_i\). The code bounds the magnitude of a negative eigenvalue relative to the largest eigenvalue by \(10^{-10}\) before removing roundoff-sized negative entries, and checks the reconstructed covariance to relative Frobenius tolerance \(10^{-9}\). The 2,200-sample ensemble is centered at each node, without rescaling its measured standard deviation to the input value. Centering preserves the usual unbiased sample-covariance estimator but weakly correlates rows of the finite ensemble. All comparisons reuse these centered realizations.

\subsection*{A common polynomial estimator and nonlinear root}
The geometry grid contains 261 points between \(2M\) and \(15M\). For the first nine points, \(x_i=r_i-r_H\) spans \(0\le x_i\le0.4M\). Let \(D_{i\alpha}=(1,x_i,x_i^2)_\alpha\), \(P=(D^{\mathsf T}D)^{-1}D^{\mathsf T}\), and
\begin{equation}
\boldsymbol c=P\delta\boldsymbol F_{\rm fit},\qquad
f(x)=c_0+c_1x+c_2x^2,\qquad
\Sigma_c=P C_{F,\rm fit}^{\rm mod}P^{\mathsf T}.
\label{eq:fitcov}
\end{equation}
The implementation uses a pseudoinverse rather than explicitly inverting \(D^{\mathsf T}D\). Both the horizon value and its derivatives are taken from this one polynomial: \(X=c_0\), \(Y=c_1\) and \(Z=2c_2\). This avoids mixing a first-node value with a separately fitted derivative. The leading covariance predictions are obtained from \(\Sigma_c\) and Eq.~\eqref{eq:widths}, rather than from numerical differentiation of a colour map.

The root-evaluated observable solves
\begin{equation}
(r_H+x)-r_H+(r_H+x)f(x)=0,\qquad
T_\xi=\frac{r_H/(r_H+x)^2+c_1+2c_2x}{4\pi}.
\label{eq:cubicroot}
\end{equation}
A realization is accepted only when it has a unique real root within \(|x|<0.35M\), a positive slope and a residual in \(F_0+f\) below \(10^{-9}\). The shared implementation stops if these conditions fail; it does not independently trim the tails of the radius and temperature samples. The polynomial also defines the continuation just inside the reference horizon. This is a reconstruction prescription, not information supplied by exterior modes about the interior quantum state. ``Root-evaluated'' means solving Eq.~\eqref{eq:cubicroot} in this model, not solving the stochastic Einstein equations exactly.

The local polynomial is used for the displayed metric samples over the fitting interval. The corresponding projected covariance is \(B C_F^{\rm mod}B^{\mathsf T}\), where \(B\) is the identity outside the fit block and equals \(DP\) within it. Histograms, paired comparisons, the amplitude scan at \(\lambda_Q=1\), and the final sampling-convergence point all use the same root and temperature arrays. The covariance panel is divided by its maximum absolute entry over the displayed window; it is a normalized covariance, not a Pearson correlation coefficient.

\subsection*{Numerical resolution and accuracy accounting}
Two numerical inputs occur in the figure set and should not be conflated. The field and source panels, Figs.~\ref{fig1}, \ref{fig2} and Supplementary Fig.~S1, use the field/source calculation: 37 field frequencies in \(0.04\le M\omega\le1.6\), \(\ell=0,\ldots,7\), 25 stress frequencies up to \(M\Omega=1.2\), and a stress integration domain \([2.05M,30M]\). The modes start at \(2M+10^{-8}M\), with matching radius \(1600M\), working precision 35 digits and requested accuracy/precision goals 18/15. These settings are not measured error bounds; the supplied residual curves remain the relevant evidence.

The shared reconstruction underlying Fig.~\ref{fig3} and Extended Data Fig.~\ref{edfig:metric_validation} uses its own higher-order source calculation, with 31 field frequencies over \([0.03,1.5]/M\), \(\ell=0,\ldots,8\), and 25 stress frequencies over \([0,1.2]/M\). Its source domain is \([2.005M,30M]\), with response nodes through \(15M\), 32-digit working precision and frequency-dependent asymptotic matching radii between \(180M\) and \(1600M\). The horizon mode start is \(2M+10^{-7}M\). The common stochastic seed is 20260930. Supplementary Information gives the source nodes, fit prescription and distinctions between source and sampling diagnostics. Supplementary Fig.~S1 measures the field/source calculation; the ensemble diagnostics address sampling and reconstruction at the separately specified geometric-source resolution.

For each figure pair, uncertainty has separate contributions from radial mode extraction, spectral and angular quadrature, radial response, smoothing and continuation, and finite sampling. The near equality of widths in two presentations of the same array checks internal consistency only. An independent validation requires refinement of the source and the reconstruction at fixed physical sampling, followed by a separate ensemble or a direct covariance prediction.

\section*{Data availability}
The data supporting the findings of this study will be provided by the corresponding author upon reasonable request.

\section*{Code availability}
The code used for the calculations and figure generation in this study will be provided by the corresponding author upon reasonable request.

\clearpage
\setcounter{figure}{0}
\renewcommand{\theHfigure}{ED\arabic{figure}}
\makeatletter
\renewcommand{\fnum@figure}{\textbf{Extended Data Fig.\ \thefigure\ |}}
\makeatother
\begin{figure*}[t]
\centering
\includegraphics[width=\textwidth]{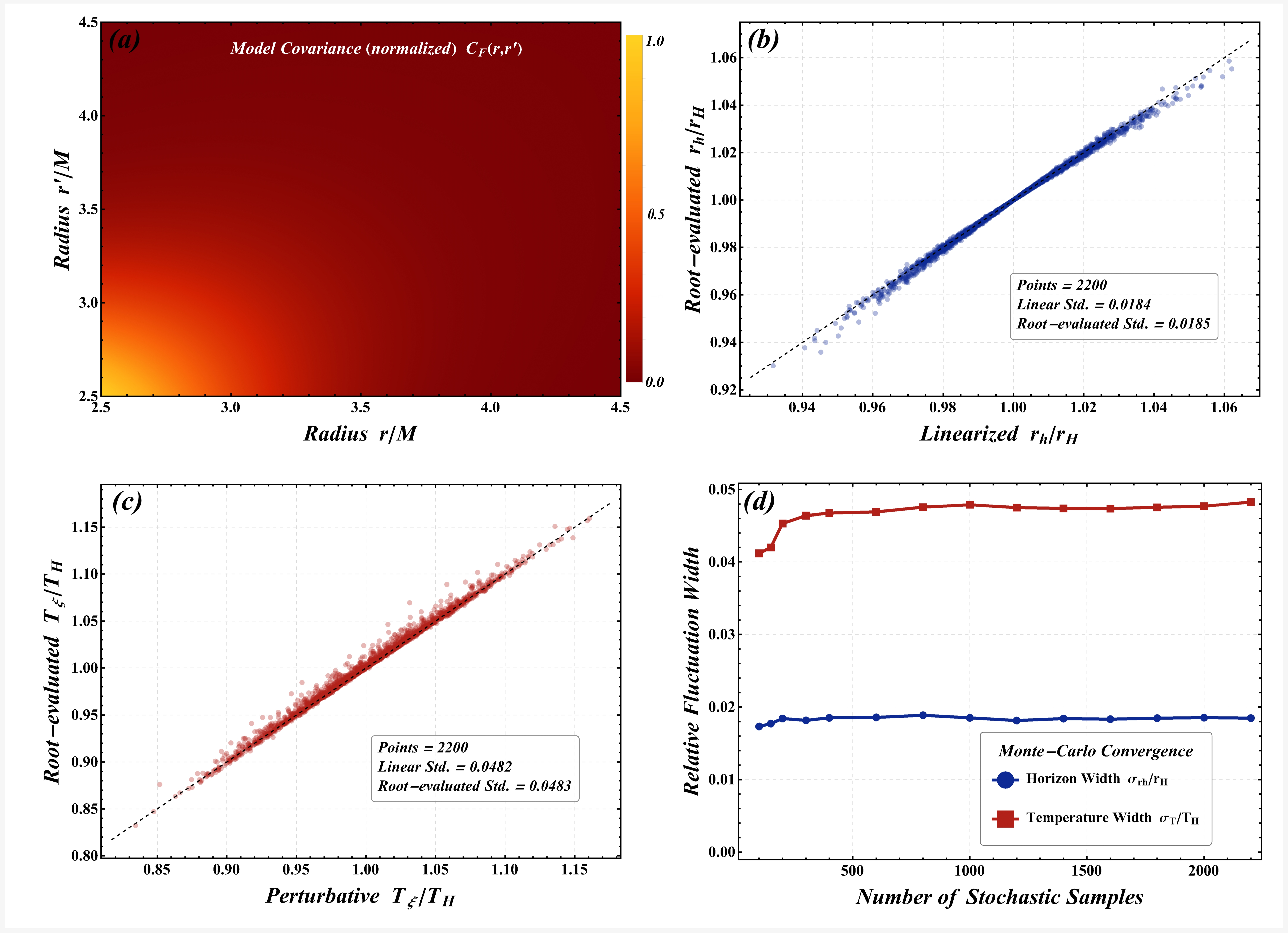}
\caption{\textbf{Spatial covariance and paired tests of the stochastic geometric reconstruction.}
All panels refer to the Gaussian lapse ensemble used in Fig.~\ref{fig3}, with 2,200 realizations, amplitude parameter \(\epsilon_F=0.018\), and common random seed 20260930. The construction uses a source-derived, smoothed RMS envelope and an imposed Gaussian radial correlation length \(L_c=0.75M\); it therefore tests a specified finite-resolution reconstruction of Hawking-field-driven geometry.
\textbf{(a)} Model lapse covariance in the radial window \(2.5\le r/M,r'/M\le4.5\). The covariance combines the radial envelope with the Gaussian correlation kernel and includes the local polynomial projection where applicable. Each entry is divided by the maximum absolute covariance in the displayed window. This normalization differs from division by the two local standard deviations and therefore does not give a Pearson correlation coefficient. The colour scale is linear from 0 (dark red) to 1 (yellow), with larger values indicating stronger covariance.
\textbf{(b)} Linearized and root-evaluated horizon radii, paired realization by realization. Both estimates use the same local lapse fit. The linear displacement is \(\delta r_h=-f(r_H)/F'_0(r_H)\); the root-evaluated estimate solves \(F_0+f=0\) within the fitted interval. The relative widths are 0.0184 and 0.0185, respectively. A realization requires a unique accepted root, a positive lapse slope and a root residual below \(10^{-9}\), as specified in Methods.
\textbf{(c)} Corresponding surface-gravity temperature parameters. The first-order change depends on \(f'(r_H)-[F''_0(r_H)/F'_0(r_H)]f(r_H)\), whereas the root-evaluated value uses the lapse slope at the displaced root. Their relative widths are 0.0482 and 0.0483. Dashed equality lines in panels \textbf{b} and \textbf{c} make the nonlinear differences visible. Solving the fitted root retains nonlinear dependence within this reconstruction; the gravitational response remains the stated reduced model.
\textbf{(d)} Width estimates obtained from nested prefixes of the same ensemble. The final root-evaluated widths, 0.0185 for radius and 0.0483 for temperature, reproduce the histograms in Fig.~\ref{fig3}. Successive points share realizations and are not independent estimates. This test assesses sampling stability at fixed covariance, fitting window and radial resolution. The paired comparisons establish that the unequal radius and temperature widths follow from their different dependence on the same fluctuations, while convergence of the microscopic source requires separate spectral, angular and radial refinement.}
\label{edfig:metric_validation}
\end{figure*}

\end{document}

\title{\texorpdfstring{Supplementary Information:\\Black Hole Horizon Fluctuations from Unruh Quantum Noise}{Supplementary Information: Black Hole Horizon Fluctuations from Unruh Quantum Noise}}
	\author{Kashif Ammar Yasir}
	\email{kayasir@zjnu.edu.cn}
	\affiliation{Department of Physics, Zhejiang Normal University, Jinhua 321004, China.}
	\affiliation{Zhejiang Institute of Photoelectronics, Jinhua 321004, China.}
	%%%%%%%%%%%%%%%%%%%%%%%%%%%%%%%%%%%%%%%%%%%%%%%%%%%%%%%%%%%%%%%%%%%%%%%%%%%%%%%%%%%%%%%%%%%%%%%%%%%%%%%%%%%%%%%%%%%%%%%%%%%
	\setlength{\parskip}{0pt}
	\setlength{\belowcaptionskip}{-10pt}

	%\pacs{42.50.Pq, 42.50.Gy, 67.85.Hj, 71.70.Ej}
	\date{1 October 2026}
	\maketitle
	
	%%%%%%%%%%%%%%%%%%%%%%%%%%%%%%%%%%%%%%%%%%%%%%%%%%%%%%%%%%%%%%%%%%%%%
	\section{Scope and conventions}
This Supplementary Information expands the mode normalization, stress contraction, stochastic response and paired geometric reconstruction. It specifies the finite-band Unruh calculation and the reduced geometry reconstruction discussed in the main text. It distinguishes three objects: the quantum field two-point function on a Schwarzschild reference, the connected stress covariance derived from it, and the covariance of a lapse obtained with a chosen radial map. Keeping these objects separate is necessary when interpreting a stochastic horizon width.

The signature is \((-+++)\), and \(c=\hbar=k_{\rm B}=1\). We retain \(G\) in analytic expressions; the plots set \(G=M=1\). Thus the plotted radial and frequency variables are \(r/M\), \(M\omega\) and \(M\Omega\). The reference lapse is \(F_0=1-2GM/r\). The field frequency \(\omega\) is positive and labels a scalar mode, whereas the stress frequency \(\Omega\) labels a quadratic observable. A signed field frequency is denoted by \(\nu\).

The reference field calculation is stationary in Schwarzschild time \(t\). Quantum averages in the Unruh state are denoted by \(\langle\cdot\rangle_U\), and averages over sampled geometry realizations by \(\langle\cdot\rangle_{\st}\). The latter are a Gaussian representation of a selected covariance, not a claim that every stress-tensor cumulant is Gaussian. The discussion of horizon observables is confined to simple roots of reconstructed spherical lapse profiles.

\section{Scalar modes and the complete exterior Unruh function}
\subsection{Radial equation and scattering normalization}
For a real massless scalar satisfying \((\Box-\zeta R)\Phi=0\), write
\begin{equation}
f^A_{\omega\ell m}(x)=\frac{e^{-i\omega t}}{\sqrt{4\pi\omega}}\frac{u^A_{\omega\ell}(r)}{r}Y_{\ell m}(\boldsymbol n),
\qquad A\in\{\mathrm{in},\mathrm{up}\}.
\end{equation}
The spherical harmonics obey \(\int\dd\Omega\,|Y_{\ell m}|^2=1\). On Schwarzschild, \(R=0\), and
\begin{equation}
\frac{\dd^2u}{\dd r_*^2}+[\omega^2-V_\ell(r)]u=0,
\quad V_\ell=F_0\left[\frac{\ell(\ell+1)}{r^2}+\frac{2GM}{r^3}\right],
\quad r_*=r+2GM\ln\left(\frac{r}{2GM}-1\right).
\end{equation}
The additive constant of \(r_*\) changes mode phases but not observables when used consistently.

Take \(h\sim e^{-i\omega r_*}\) at the horizon and \(h\sim A_{\rm in}e^{-i\omega r_*}+A_{\rm out}e^{i\omega r_*}\) at infinity. Its conserved current is
\begin{equation}
j[h]=\frac{h^*\partial_{r_*}h-h\partial_{r_*}h^*}{2i}=-\omega,
\qquad |A_{\rm in}|^2-|A_{\rm out}|^2=1.
\end{equation}
It follows that
\begin{align}
u^{\rm in}&=h/A_{\rm in},\\
u^{\rm up}&=h^*-(A_{\rm out}^*/A_{\rm in})h.
\label{eq:supup}
\end{align}
Near the horizon, \(u^{\rm up}\) has unit coefficient of \(e^{+i\omega r_*}\). At infinity, its ingoing coefficient vanishes and its outgoing coefficient is \(1/A_{\rm in}\). Multiplying Eq.~\eqref{eq:supup} by \(A_{\rm in}\) would change the incident normalization and would require a compensating change in the mode-sum weight.

At finite matching radius, the plane waves must be supplemented by asymptotic corrections. Increasing numerical working precision alone does not remove an error from an inadequate matching basis or an unstable fit window. Transmission and reflection are obtained independently as \(\Gamma=|A_{\rm in}|^{-2}\) and \(\mathcal R=|A_{\rm out}/A_{\rm in}|^2\). Their sum tests extraction as well as radial integration.

\subsection{Finite-radius asymptotic extraction}
For an asymptotic sign \(\varsigma=\pm1\), write
\begin{equation}
b_{\varsigma,p}(r)=e^{i\varsigma\omega r_*}\sum_{n=0}^{p}a_n^{(\varsigma)}r^{-n},\qquad a_0^{(\varsigma)}=1.
\end{equation}
Substitution into the radial equation gives, for the amplitude \(A(r)\),
\begin{equation}
F_0 A''+(F_0'+2i\varsigma\omega)A'
 -\left[\frac{\ell(\ell+1)}{r^2}+\frac{r_H}{r^3}\right]A=0.
\end{equation}
Equating coefficients of \(r^{-m-1}\) yields
\begin{equation}
a_m^{(\varsigma)}=
\frac{[m(m-1)-\ell(\ell+1)]a_{m-1}^{(\varsigma)}
-r_H(m-1)^2a_{m-2}^{(\varsigma)}}{2i\varsigma\omega m},
\qquad a_{-1}^{(\varsigma)}=0.
\label{eq:asym_rec}
\end{equation}
Both the value and the tortoise derivative are used in matching:
\begin{equation}
\partial_{r_*}b_{\varsigma,p}=e^{i\varsigma\omega r_*}
\left[i\varsigma\omega\sum_{n=0}^{p}a_n^{(\varsigma)}r^{-n}
-F_0\sum_{n=1}^{p}n a_n^{(\varsigma)}r^{-n-1}\right].
\end{equation}
At the extraction nodes, the two columns of the design matrix are \((b_-,\partial_{r_*}b_-/\omega)\) and \((b_+,\partial_{r_*}b_+/\omega)\). Least squares matches them to \((h,\partial_{r_*}h/\omega)\). Dividing derivative rows by \(\omega\) puts both types of row on comparable scales. The normalized residual is \(\|\boldsymbol y-\mathsf A\boldsymbol\alpha\|/\|\boldsymbol y\|\), with \(\boldsymbol\alpha=(A_{\rm in},A_{\rm out})\).

The shared reconstruction uses a nominal order \(p=\max(12,\ell+8)\) and compares neighboring choices of order as specified by the executable. This is an asymptotic expansion, not a convergent series at arbitrary radius; stability under radius and order variation is required. The up-mode mixing coefficient \(-A_{\rm out}^*/A_{\rm in}\), transmitted flux and fit residual are distinct diagnostics. In an opaque channel, a current formed by subtracting large nearly equal quantities can be ill-conditioned. A small relative least-squares residual alone is also insufficient to guarantee accurate reflection phases or derivatives.

\subsection{Three branches and their identities}
Let \(q^A_{\omega\ell}=u^A_{\omega\ell}/r\), \(\tau=t-t'\), \(\mu=\boldsymbol n\cdot\boldsymbol n'\), and \(p_\ell(\omega)=(2\ell+1)/(16\pi^2\omega)\). The angular addition theorem gives
\begin{align}
G_U^+(\tau;r,r',\mu)=\sum_\ell\int_0^\infty\dd\omega\,p_\ell P_\ell(\mu)
\big\{&e^{-i\omega\tau}[q^{\rm in}(r)q^{{\rm in}*}(r')+(1+n_H)q^{\rm up}(r)q^{{\rm up}*}(r')]\nonumber\\
&+e^{+i\omega\tau}n_H q^{{\rm up}*}(r)q^{\rm up}(r')\big\}.
\label{eq:supWU}
\end{align}
Here \(n_H=(e^{8\pi GM\omega}-1)^{-1}\). The incoming sector is a vacuum sector. The outgoing sector has the spontaneous and stimulated positive-frequency weight \(1+n_H\), and the negative-frequency weight \(n_H\) \cite{Unruh1976,Hodgkinson2014}. The latter multiplies a conjugated spatial mode, not merely the same positive-frequency radial product with a different coefficient.

The function satisfies \(G_U^+(x,x')^*=G_U^+(x',x)\). At fixed positive \(\omega\), the up-sector negative branch is \(e^{-\omega/T_H}\) times the complex conjugate of the up-sector positive branch. This is an up-sector relation, not global thermal detailed balance for the full Unruh state. In particular, a thermal factor must not be assigned to the incoming vacuum sector.

The equal-time shell used for a field-coherence diagnostic is the sum of all three spatial terms in Eq.~\eqref{eq:supWU}. If that shell is denoted by \(K_\omega(r,r')\), its amplitude coherence is
\begin{equation}
\mathcal C_\Phi(\omega;r,r')=
\frac{|K_\omega(r,r')|}{[K_\omega(r,r)K_\omega(r',r')]^{1/2}}.
\end{equation}
A gradient coherence replaces both fields with the same chosen radial derivative. Positivity gives \(0\le\mathcal C\le1\), provided the diagonal entries are positive. The equal-time shell contains both signs of frequency and is not the positive-frequency Wightman spectrum alone. The radial derivative must act on \(u/r\), including its \(-u/r^2\) contribution.

\section{Connected stress noise}
\subsection{Curvature coupling and a freely falling density}
Although the radial equation does not depend on \(\zeta\) on this background, the stress tensor does. With the convention \((\Box-\zeta R)\Phi=0\), its classical form is
\begin{align}
T_{\mu\nu}={}&(1-2\zeta)\nabla_\mu\Phi\nabla_\nu\Phi
 +(2\zeta-\tfrac12)g_{\mu\nu}(\nabla\Phi)^2
 -2\zeta\Phi\nabla_\mu\nabla_\nu\Phi\nonumber\\
&+2\zeta g_{\mu\nu}\Phi\Box\Phi+\zeta G_{\mu\nu}\Phi^2.
\end{align}
The quantum operator is defined by point splitting and symmetric ordering. The Ricci-flat, on-shell form is adequate for the separated-point, finite-band construction here; it is not a prescription for removing the local Hadamard singularity of an unrestricted coincidence limit \cite{BirrellDavies1982,PhillipsHu2001}.

For free fall from rest at infinity, set \(v_r=\sqrt{2GM/r}\). One may use
\begin{equation}
D_u=F_0^{-1}\partial_t-v_r\partial_r,
\qquad D_s=-v_rF_0^{-1}\partial_t+\partial_r,
\qquad D_\theta=r^{-1}\partial_\theta,\quad
D_\varphi=(r\sin\theta)^{-1}\partial_\varphi.
\end{equation}
Then \(\rho=T_{\mu\nu}u^\mu u^\nu\) reduces to
\begin{equation}
\rho=\tfrac12(D_u\Phi)^2+(\tfrac12-2\zeta)[(D_s\Phi)^2+(D_\theta\Phi)^2+(D_\varphi\Phi)^2]-2\zeta\Phi D_u^2\Phi.
\end{equation}
The calculation takes \(\zeta=1/6\). Coefficients in \(D_u\) depend on radius, so \(D_u^2\) differentiates them as well as the mode. For a negative-frequency up mode, both \(\partial_t\to+i\omega\) and spatial conjugation are required. Angular derivatives must be taken before coincidence or angular projection.

\subsection{Explicit mode derivatives in the falling frame}
For a signed-frequency factor \(e^{-i\nu t}q(r)\), define \(F=F_0\) and \(v_r=\sqrt{r_H/r}\). The nonangular derivatives reduce to
\begin{align}
D_u q&=-\frac{i\nu}{F}q-v_rq',&
D_s q&=\frac{i\nu v_r}{F}q+q',\\
D_u^2q&=v_r^2q''+
\left(v_rv_r'+\frac{2i\nu v_r}{F}\right)q'
+\left(-\frac{\nu^2}{F^2}-\frac{i\nu v_rF'}{F^2}\right)q,
\qquad v_r'=-\frac{v_r}{2r}.
\label{eq:Du2}
\end{align}
Here the notation means the radial coefficient after the time factor is removed. The radial field equation also gives
\begin{equation}
q''=-\left(\frac{2}{r}+\frac{F'}{F}\right)q'
-\left(\frac{\nu^2}{F^2}-\frac{\ell(\ell+1)}{r^2F}\right)q.
\end{equation}
Thus second radial derivatives can be evaluated from the ODE instead of differentiating a sampled interpolation twice. For the negative-frequency up branch, take \(\nu=-\omega\) and \(q=q^{{\rm up}*}\) together. Merely changing the sign of \(\nu\), or conjugating only the radial function, is not the negative-frequency mode.

Angular derivatives act on the addition theorem
\begin{equation}
\sum_mY_{\ell m}(\boldsymbol n)Y_{\ell m}^*(\boldsymbol n')
=\frac{2\ell+1}{4\pi}P_\ell(\mu).
\end{equation}
For example,
\(\nabla_A\nabla_{B'}P_\ell(\mu)
=P_\ell''(\mu)(\nabla_A\mu)(\nabla_{B'}\mu)
+P_\ell'(\mu)\nabla_A\nabla_{B'}\mu\).
The factors \(1/r\) and \(1/r'\) converting coordinate derivatives into orthonormal components are applied separately. Taking angular coincidence before these derivatives would incorrectly delete angular stresses.

\subsection{Wick contraction and symmetrization}
For the six operators \(X=(\Phi,D_u\Phi,D_s\Phi,D_\theta\Phi,D_\varphi\Phi,D_u^2\Phi)\), write \(\rho=C_{ab}X_aX_b\), with the coefficient matrix specified in Methods. Define \(W_{ac}(x,y)=\langle X_a(x)X_c(y)\rangle_U\). The disconnected pairing is removed by subtracting the mean density; the two remaining pairings give
\begin{equation}
C_\rho^>(x,y)=C_{ab}C_{cd}(W_{ac}W_{bd}+W_{ad}W_{bc}).
\end{equation}
For symmetric \(C\), a bilinear contraction of two derivative matrices can be evaluated as
\begin{equation}
\mathcal B(W_1,W_2)=2\Tr(CW_1CW_2^{\mathsf T}).
\end{equation}
The transpose here is not a Hermitian conjugate. Complex conjugation is supplied by the mode products and, subsequently, by the required exchange of spacetime arguments.

The symmetrized noise is \(N_\rho(x,y)=\tfrac12[C_\rho^>(x,y)+C_\rho^>(y,x)]\), equivalently the real part at fixed spacetime points. If \(W(\tau)=\int\dd\nu\,e^{-i\nu\tau}\mathcal W(\nu)\), then
\begin{equation}
\widetilde C^>_{ij}(\Omega)=2\pi\int_{-\infty}^{\infty}\dd\nu\,
\mathcal B[\mathcal W_{ij}(\nu),\mathcal W_{ij}(\Omega-\nu)],
\quad
\widetilde N_{ij}(\Omega)=\tfrac12[\widetilde C^>_{ij}(\Omega)+\widetilde C^>_{ji}(-\Omega)].
\label{eq:supconv}
\end{equation}
The final expression is Hermitian in its spatial indices. Its diagonal is real and nonnegative for a correctly constructed autocorrelation. Entrywise reality is not required off the diagonal. The covariance of any real weighted source is nonnegative, which provides a stronger test than checking only the diagonal entries.

For \(\mathcal I_+=[\omega_{\min},\omega_{\max}]\) and \(\mathcal I_-=[-\omega_{\max},-\omega_{\min}]\), write the spectrum as its two supported branches. Each contraction has the integration domain
\begin{equation}
\mathcal I_{st}(\Omega)=\mathcal I_s\cap(\Omega-\mathcal I_t),\qquad
\widetilde C^>(\Omega)=2\pi\sum_{s,t\in\{+,-\}}\int_{\mathcal I_{st}(\Omega)}\!\dd\nu\,
\mathcal B[\mathcal W_s(\nu),\mathcal W_t(\Omega-\nu)].
\end{equation}
The positive branch includes both incoming and outgoing terms before products are formed. The mixed-sign intervals generate difference frequencies and generally contribute at \(\Omega=0\). Setting the negative branch to zero would remove these terms. Interpolation is confined to each supported interval; bridging the excluded central gap would add uncomputed field modes.

No vacuum subtraction is made in Eq.~\eqref{eq:supconv}. If an excess quantity relative to a reference state were desired, it would require \(N_U-N_{\rm ref}\), not a Wick contraction of \(G_U^+-G_{\rm ref}^+\). The difference of two positive covariance kernels is not generally positive and cannot automatically be used as the covariance of a classical random source.

\subsection{Angular monopole of a quadratic observable}
The monopole density is \(\rho_0(t,r)=(4\pi)^{-1}\int\dd\Omega_{\boldsymbol n}\rho(t,r,\boldsymbol n)\). Its spectrum is
\begin{equation}
\widetilde N_0(\Omega;r,r')=\frac{1}{(4\pi)^2}\int\dd\Omega_{\boldsymbol n}\dd\Omega_{\boldsymbol n'}\,
\widetilde N_\rho(\Omega;r,r',\boldsymbol n\cdot\boldsymbol n')
=\frac12\int_{-1}^{1}\dd\mu\,\widetilde N_\rho(\Omega;r,r',\mu).
\end{equation}
All retained scalar multipoles enter the stress contraction before this integral. A stress monopole can be formed from products of higher scalar multipoles. Thus the spherical source calculation does not reduce to the scalar \(\ell=0\) mode. The local density spectrum instead sets \(\mu=1\), and cannot be interchanged with \(N_0\).

\section{Open-system derivation and the status of the Langevin equation}
A matter field coupled to a metric perturbation has the linear interaction \(\delta S_\Phi=\tfrac12\int\sqrt{-g}\,h_{\mu\nu}T^{\mu\nu}\). Defining the source \(J=h/2\) fixes otherwise ambiguous factors of two in the influence functional. For forward and backward sources, \(J_\Delta=J^+-J^-\) and \(J_c=(J^++J^-)/2\), the connected second-cumulant expansion reads
\begin{equation}
S_{\rm IF}=\int J_\Delta\langle T\rangle_U
 +\int J_\Delta\chi_{\rm ret}J_c
 +\frac{i}{2}\int J_\Delta N J_\Delta+O(J^3).
\label{eq:supIF}
\end{equation}
Invariant integration measures and tensor indices are implicit. With action coupling \(+\int JT\), the nonlocal linear susceptibility is \(\chi_{\rm ret}(x,x')=i\Theta(t-t')\langle[\hat T(x),\hat T(x')]\rangle\). Local contact terms from the metric dependence of the stress operator belong to the complete response as well. The imaginary part contains the anticommutator noise, whereas the causal response contains the commutator. They cannot be substituted for one another.

For every real test tensor \(f\), the smeared noise obeys
\begin{equation}
\int f N f=\langle\hat t[f]^2\rangle_U\ge0.
\end{equation}
A positive kernel therefore admits a Gaussian characteristic functional,
\begin{equation}
\langle e^{i\int f\xi}\rangle_{\st}=e^{-\frac12\int fNf}.
\end{equation}
Applying this identity to Eq.~\eqref{eq:supIF} and varying the real stochastic action gives the Einstein--Langevin equation at linear order in \(h\). If \(\mathcal L_{\rm ret}\) denotes its complete linear operator and \(G_{\rm ret}\) its causal inverse with specified initial/boundary conditions, the induced solution and covariance are
\begin{align}
h&=h_{\rm hom}+8\pi G\,G_{\rm ret}*\xi,\\
\langle h h\rangle_{\st,c}&=C_{\rm hom}+(8\pi G)^2G_{\rm ret}*N*G_{\rm ret}^{\mathsf T},
\label{eq:full_response}
\end{align}
assuming independent homogeneous initial fluctuations and the source. A general initial state can also contain cross-correlations. The radial integral used in this paper is a restricted map in place of this full causal response; its simpler kernel is not a computed \(G_{\rm ret}\).

The influence-functional expansion follows nonequilibrium field theory \cite{FeynmanVernon1963,CalzettaHu2008}. This derivation is established in stochastic semiclassical gravity \cite{HuMatacz1995,MartinVerdaguer1999,HuVerdaguer2008}, including black-hole applications \cite{Sinha2003,HuRoura2007}. The present construction specifies the complete retained-band Unruh input and the successive projections and reconstruction used in the figures. Moreover, an Einstein--Langevin equation with a classical stochastic source reproduces selected symmetrized quantum correlations at its stated order; it does not assert a fully quantized nonlinear metric dynamics. The Unruh state is not a global KMS state, so a single equilibrium \(\coth(\Omega/2T_H)\) factor is not used to infer its complete noise from dissipation.

\section{From density covariance to a reduced lapse covariance}
\subsection{Exact mass equation and the closure used here}
For an ingoing spherical metric
\begin{equation}
\dd s^2=-e^{2\psi}F\dd v^2+2e^\psi\dd v\dd r+r^2\dd\Omega^2,
\qquad F=1-\frac{2Gm}{r},
\end{equation}
the radial mass equation is \(\partial_r m=-4\pi r^2T^v{}_v\). Other equations determine the mass flux and the radial redshift function. In general, \(-T^v{}_v\ne T_{\mu\nu}u^\mu u^\nu\). Replacing the former by a free-fall density while fixing \(\psi=0\) therefore defines a restricted model. It is not an exact projection of all tensor Einstein equations for the Unruh source.

Let \(\eta\) have the monopole free-fall density covariance above. The displayed model adopts
\begin{equation}
\partial_r\delta m=4\pi r^2\eta,\qquad \delta m(R_B)=0,
\qquad \delta F(r)=\frac{8\pi G}{r}\int_r^{R_B}\dd s\,s^2\eta(s).
\label{eq:supclosure}
\end{equation}
Its response has no propagation delay. By contrast, a full linear stochastic semiclassical calculation includes a retarded stress-polarization kernel, tensor conservation and the redshift sector \cite{MartinVerdaguer1999,HuVerdaguer2008}. The finite-radius condition in Eq.~\eqref{eq:supclosure} fixes the outer mass, not an asymptotic ADM completion by itself. The plots also do not include a separately computed self-consistent mean stress correction to Schwarzschild.

\subsection{Time slicing of a bilocal spectrum}
The mode spectrum is defined in \(t\), whereas the mass equation above is naturally written on an ingoing \(v=t+r_*\) slice. At equal \(v\), two different radii have \(t-t'=-(r_*-r_*')\). For the stated Fourier convention, the transformation of a scalar-projected bilocal spectrum is
\begin{equation}
\widetilde N^{(v)}(\Omega;r,r')=
 e^{i\Omega[r_*(r)-r_*(r')]}\widetilde N^{(t)}(\Omega;r,r').
\end{equation}
There is no such phase on the radial diagonal. It matters in a double radial integral. A tensor component transformation is additionally needed if the source is changed from free-fall density to \(-T^v{}_v\). A radial map applied directly to the unshifted \(t\)-spectrum should be interpreted on that chosen slicing, rather than as an exact equal-\(v\) constraint solution.

\subsection{Covariance and radial quadrature}
Set \(K(r,s)=8\pi Gs^2\Theta(s-r)\Theta(R_B-s)/r\). Then
\begin{equation}
S_F(\Omega;r,r')=\int\dd s\dd s'\,K(r,s)\widetilde N_0(\Omega;s,s')K(r',s'),
\qquad
C_F(r,r')=\int_0^{\Omega_{\max}}\frac{\dd\Omega}{\pi}\Re S_F(\Omega;r,r').
\end{equation}
The real part is required for the equal-time covariance of real fields. Dropping off-diagonal source entries changes the model. The equal-time covariance is sufficient for drawing spatial realizations, but does not specify a causal time evolution of those realizations.

For source values \(\eta_i\) at nodes \(s_i\), with \(C_{0,ij}=\langle\eta_i\eta_j\rangle\), write \(\delta F_a=\sum_iK_{ai}\eta_i\). Each \(K_{ai}\) includes its radial quadrature weight, and \(C_F=KC_0K^{\mathsf T}\). Weights must not be inserted a second time into \(C_0\). The shell mass variance is \((4\pi s_i^2w_i)^2C_{0,ii}\), which depends on the discretization width. A cumulative diagnostic uses \(k_i(R)=8\pi Gs_i^2w_i/r_H\) on \([r_{\min},R]\) and has variance \(k(R)C_0k(R)^{\mathsf T}\). It need not be monotone in \(R\), because covariance terms from added shells can have either sign.

\section{Finite resolution and the field diagnostics}
The field/source implementation associated with main Figs.~1--2 and Supplementary Fig.~S1 uses the parameters in Table~\ref{tab:resolution}. The 37 frequencies come from joining three subgrids, including shared endpoints only once. The signed spectrum is supported on \([-\omega_{\max},-\omega_{\min}]\cup[\omega_{\min},\omega_{\max}]\). Convolution intervals are the intersections of this support with its shift by \(\Omega\); no interpolation is performed across the central excluded interval. Piecewise-linear spectral interpolation and trapezoidal integration remain numerical approximations that require refinement tests.

\begin{table}[h]
\caption{Resolution of the associated finite-band source calculation. These are input settings, not measured accuracy guarantees.}
\label{tab:resolution}
\begin{tabular}{ll}
\hline\hline
Quantity & Setting\\
\hline
Scalar coupling & \(\zeta=1/6\)\\
Scalar multipoles & \(\ell=0,1,\ldots,7\)\\
Field-frequency grids & 13 points on \([0.04,0.2]\), 15 on \([0.2,0.8]\), 11 on \([0.8,1.6]\)\\
Distinct field frequencies & 37, in units of \(M^{-1}\)\\
Stress-frequency grid & 25 points on \([0,1.2]\), in units of \(M^{-1}\)\\
Radial mode start and matching radii & \(2M+10^{-8}M\), \(1600M\)\\
Stress integration domain & \([2.05M,30M]\)\\
Displayed source/response domain & \([2.05M,10M]\)\\
Working precision & 35 decimal digits\\
Requested accuracy/precision goals & 18/15\\
\hline\hline
\end{tabular}
\end{table}

The radial source nodes, in units of \(M\), are
\begin{equation}
\{2.05,2.10,2.15,2.25,2.40,2.60,2.90,3.30,3.80,4.50,5.50,6.80,8.50,10,13,17,22,30\}.
\end{equation}
The field band would permit stress sum frequencies up to \(M\Omega=3.2\); the reported variance uses only the plotted stress band up to 1.2. That additional restriction is part of the definition of the reported observables. The numerical start of a radial mode near \(2M\) does not imply that the stress covariance has been sampled all the way to the horizon. A sharp concentration in a truncated, coincident spectrum is also not evidence that the Unruh state itself is singular at the future horizon. State regularity and the existence of an unsmeared stress coincidence limit are different questions.

\begin{figure*}[t]
\centering
\includegraphics[width=\textwidth]{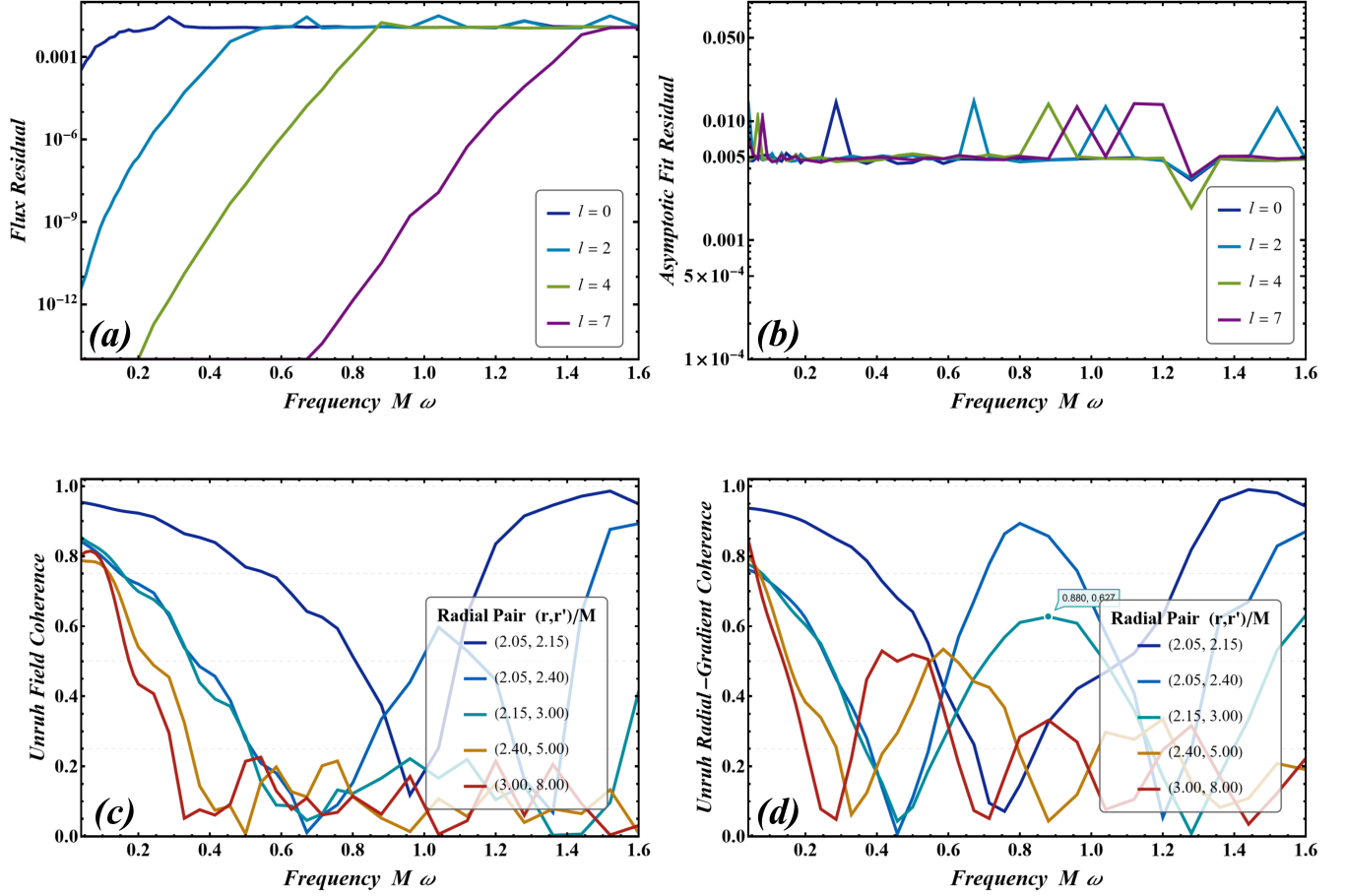}
\caption{\textbf{Scattering residuals and finite-band Unruh coherences.}
\textbf{(a)} Flux residual for selected scalar multipoles. Values reach the percent scale in the displayed range.
\textbf{(b)} Asymptotic-fit residuals, with typical values of several \(10^{-3}\) and larger excursions. The displayed curves do not constitute an independent up-mode leakage measurement.
\textbf{(c,d)} Field and radial-gradient amplitude coherences for the listed radial pairs. The minima and revivals reflect mode interference within the retained band. Bounds between zero and one are consistency checks, not convergence tests. The isolated coordinate tooltip in panel d is a plotting annotation, not an additional datum.}
\label{sfig:mode_validation}
\end{figure*}

Figure~\ref{sfig:mode_validation} replaces claims of high-precision validation by the residuals actually displayed. A small current drift of the raw ODE solution does not guarantee stable asymptotic coefficients: it can coexist with a much larger extraction error. Useful independent checks include matching-window drift, asymptotic-order variation, transmitted/reflected flux balance and stability of the up-mode combination. The stress tensor further uses derivatives, so its error may be more sensitive than a field amplitude or luminosity.

The luminosity spectrum is \(\sum_\ell(2\ell+1)\omega\Gamma_{\omega\ell}n_H/(2\pi)\). Its angular remainder is strongly thermally suppressed. Vacuum pieces of Eq.~\eqref{eq:supWU} do not have that suppression. Convergence of the radiative tail consequently does not establish angular or ultraviolet convergence of the stress noise. A hard mode cutoff gives a finite model; a continuum statement requires a specified spacetime sampling function and control of the associated renormalized distribution \cite{HuRoura2007}.

\subsection{The separate source calculation for the shared geometry ensemble}
Main Fig.~3 and Extended Data Fig.~1 use a common executable source calculation with 31 uniformly spaced field frequencies, \(0.03\le M\omega\le1.5\), and nine scalar multipoles, \(\ell=0,\ldots,8\). Its 25 stress frequencies cover \(0\le M\Omega\le1.2\). The source nodes are
\begin{equation}
\begin{split}
\{&2.005,2.020,2.050,2.100,2.180,2.300,2.480,2.700,3.000,\\
 &3.400,4.000,5.000,6.500,8.500,11.000,15.000,22.000,30.000\}M.
\end{split}
\end{equation}
Response profiles use the subset through \(15M\), with the outer mass fixed at \(30M\). The mode integration begins at \(2M+10^{-7}M\), uses 32-digit working precision and selects matching radii in \([180M,1600M]\). Eighteen extraction nodes constrain the asymptotic value/derivative fit. The declared checks are fit and flux residuals below \(5\times10^{-4}\), order sensitivity below \(2\times10^{-4}\), and up-mixing sensitivity below \(5\times10^{-4}\). They are acceptance diagnostics to inspect in the run log, not achieved values inferred from the EPS files. The code permits diagnostic output when these source checks warn; a plotted reconstruction alone does not imply that they passed.

The direct response covariance has a separate negative-eigenvalue diagnostic, with a relative warning threshold \(5\times10^{-3}\). Its diagonal is used for the reconstruction envelope only after a finiteness and negative-variance check. The positive Gaussian model used subsequently must not conceal an indefinite direct response covariance. Raw source arrays and diagnostics must be retained to test that issue independently.

The mathematical branch construction is common to the field/source panels and the shared geometry calculation, but their numerical parameters are not identical. Supplementary Fig.~S1 belongs to the former. It cannot be relabelled as a convergence test of the latter. Comparisons across the two groups are consequently structural and qualitative until their source grids and diagnostics are matched.

\section{Common Gaussian reconstruction for the geometric figures}
\subsection{Separating the microscopic response from the sampling model}
Write \(C_F^{\rm rad}\) for the finite-band radial response covariance. Its diagonal gives \(s^{\rm raw}(r)=\sqrt{C_F^{\rm rad}(r,r)}\). For the geometric ensemble, this profile is linearly interpolated and anchored at \(r_H\) by its first resolved exterior value. On the uniform grid \(r_i/M=2+0.05(i-1)\), \(i=1,\ldots,261\), the reconstruction applies
\begin{equation}
\bar s_i=\frac{\sum_j w_j e^{-(r_i-r_j)^2/(2L_s^2)}s_j^{\rm raw}}
 {\sum_j w_j e^{-(r_i-r_j)^2/(2L_s^2)}},\qquad
s_i=\epsilon_F\bar s_i/\bar s_1,
\end{equation}
where \(w_j\) are trapezoidal weights. The sampling covariance is
\begin{equation}
C_{ij}^{\rm mod}=\lambda_Q^2s_i s_j R_{ij},\qquad
R_{ij}=\exp[-(r_i-r_j)^2/(2L_c^2)],\quad
L_s=L_c=0.75M,\quad\epsilon_F=0.018.
\label{eq:supmodel}
\end{equation}
The Gaussian kernel \(R\) is positive semidefinite. Multiplication on both sides by the diagonal amplitude matrix preserves that property. However, Eq.~\eqref{eq:supmodel} replaces the off-diagonal structure of \(C_F^{\rm rad}\); it is not a direct factorization of the microscopic covariance. Its normalization also removes the absolute scale of the raw RMS. Consequently the percentage widths cannot be read as parameter-free predictions for a black hole of specified physical mass.

Envelope smoothing differs from field smearing. If a random field itself were linearly averaged by a matrix \(A\), its covariance would be \(A C A^{\mathsf T}\). Here \(A\) first acts on the diagonal RMS envelope and a separate correlation kernel is imposed afterwards. This distinction is essential for interpreting the covariance map and its derivatives. The term ``complete Unruh'' describes retention of all state branches in the quantum input, not removal of every approximation in the gravitational reconstruction.

\subsection{Sampling, centering and paired comparisons}
Diagonalize \(C^{\rm mod}=Q\Lambda Q^{\mathsf T}\) and generate independent rows \(\boldsymbol z^{(n)}\) of unit normal variables. The samples are \(\delta\boldsymbol F^{(n)}=Q\Lambda^{1/2}\boldsymbol z^{(n)}\). Only negative eigenvalues bounded by \(10^{-10}\max|\Lambda|\) are eligible for roundoff removal; the relative covariance reconstruction error must be below \(10^{-9}\). These checks concern the sampling covariance, separately from the microscopic response spectrum.

At \(N_s=2200\) and seed 20260930, the code centers each column by subtracting its sample mean, without multiplying the realizations to impose an exact observed RMS. For originally independent Gaussian samples, centering gives
\begin{equation}
\langle\delta F_i^{(n)}\delta F_j^{(m)}\rangle
=\left(\delta_{nm}-\frac1{N_s}\right)C_{ij}^{\rm mod}.
\end{equation}
Thus a single centered row has covariance \((1-1/N_s)C^{\rm mod}\), while the usual sample covariance with denominator \(N_s-1\) remains unbiased. The displayed covariance and measured finite-sample widths should not be forced into equality by an additional rescaling.

The amplitude scan uses \(\lambda_Q\delta F^{(n)}\) for the same underlying rows. It includes \(\lambda_Q=1\) explicitly. Histograms, the root-evaluated axes of the comparison plots, and the final convergence entries use identical observable arrays. Sampling curves use nested prefixes and hence their plotted points are not statistically independent. Agreement between duplicate representations of an array is a useful implementation check but not an independent source calculation.

\subsection{Finite-window value and derivative estimation}
Let \(\boldsymbol f\) contain the first nine field samples, with \(x_i=r_i-r_H\in[0,0.4M]\). The least-squares polynomial and its covariance follow from
\begin{align}
D_{i\alpha}&=(1,x_i,x_i^2)_\alpha,&
P&=(D^{\mathsf T}D)^{-1}D^{\mathsf T},\\
\boldsymbol c&=P\boldsymbol f,&
\Sigma_c&=P C_{\rm fit}^{\rm mod}P^{\mathsf T}.
\end{align}
A pseudoinverse is used numerically. The constant and linear coefficients estimate \(X=\delta F(r_H)\) and \(Y=\delta F'(r_H)\), and \(Z=\delta F''(r_H)=2c_2\). Defining \(a=F_0'(r_H)\) and \(b=F_0''(r_H)\), the normalized linear observables are
\begin{equation}
\begin{pmatrix}\delta r_h/r_H\\ \delta T/T_H\end{pmatrix}
=L\boldsymbol c,\qquad
L=\begin{pmatrix}-1/(a r_H)&0&0\\-b/(4\pi aT_H)&1/(4\pi T_H)&0\end{pmatrix},
\qquad \Sigma_{rT}=L\Sigma_cL^{\mathsf T}.
\label{eq:linear_matrix}
\end{equation}
This is the finite-grid counterpart of differentiating a covariance at coincidence. It makes the fitting window part of the observable definition. A derivative fitted over a different physical interval need not have the same variance even when both procedures use nine points.

The metric display uses the projected values \(DP\boldsymbol f\) in the local interval and the original samples elsewhere. With this projection embedded in the full matrix \(B\), its covariance is \(B C^{\rm mod}B^{\mathsf T}\). The map in Extended Data Fig.~1 displays this model covariance divided by the largest absolute entry in the plotted subdomain \([2.5M,4.5M]^2\). That normalization is different from the correlation coefficient \(C(r,r')/\sqrt{C(r,r)C(r',r')}\).

\subsection{Nonlinear roots, slopes and mean shifts}
For a simple reference root, expand \(F_\epsilon=\bar F+\epsilon f\) and \(r_h=\bar r_h+\epsilon r_1+\epsilon^2r_2+\cdots\). Set \(a=\bar F'\), \(b=\bar F''\), \(c=\bar F'''\), and \(X=f\), \(Y=f'\), \(Z=f''\) at that root. The root equation gives successively
\begin{equation}
a r_1+X=0,\qquad a r_2+\tfrac12 b r_1^2+Yr_1=0,
\end{equation}
so that
\begin{equation}
r_1=-X/a,\qquad r_2=XY/a^2-bX^2/(2a^3).
\end{equation}
The surface-gravity parameter obeys
\begin{align}
4\pi T_\epsilon={}&a+\epsilon(Y-bX/a)\nonumber\\
&+\epsilon^2\left[\frac{bXY}{a^2}
 +\left(\frac{c}{2a^2}-\frac{b^2}{2a^3}\right)X^2-\frac{XZ}{a}\right]+O(\epsilon^3).
\label{eq:secondorderT}
\end{align}
These are expansions of the root and slope in a prescribed random profile, not a second-order gravitational field equation. Centered \(X,Y,Z\) can generate nonzero second-order means. For example,
\begin{align}
\langle r_h\rangle-\bar r_h&=\epsilon^2\left[\frac{\langle XY\rangle}{a^2}-\frac{b\langle X^2\rangle}{2a^3}\right]+O(\epsilon^3),\\
4\pi(\langle T\rangle-T_0)&=\epsilon^2\left[\frac{b\langle XY\rangle}{a^2}
 +\left(\frac{c}{2a^2}-\frac{b^2}{2a^3}\right)\langle X^2\rangle-\frac{\langle XZ\rangle}{a}\right]+O(\epsilon^3).
\end{align}
For Schwarzschild, \(a=1/r_H\), \(b=-2/r_H^2\), \(c=6/r_H^3\), and these expansions simplify to
\begin{align}
\frac{r_h}{r_H}&=1-\epsilon X+\epsilon^2(X^2+r_HXY)+O(\epsilon^3),\\
\frac{T}{T_H}&=1+\epsilon(2X+r_HY)
 +\epsilon^2(X^2-2r_HXY-r_H^2XZ)+O(\epsilon^3).
\end{align}
For centered Gaussian coefficients, odd moments vanish, and the leading correction to the linear variance is of order \(\epsilon^4\). The absolute correction to a nonzero standard deviation is therefore generally order \(|\epsilon|^3\), although the pointwise root/linear difference begins at order \(\epsilon^2\).

In the implementation, \(f(x)=c_0+c_1x+c_2x^2\), and the nonlinear root is found from the cubic
\begin{equation}
c_2x^3+(c_1+r_Hc_2)x^2+(1+c_0+r_Hc_1)x+r_Hc_0=0.
\end{equation}
The code requires exactly one real candidate in \(|x|<0.35M\), positive \(T\), and \(|F_0(r_H+x)+f(x)|<10^{-9}\). The same polynomial supplies \(T=[r_H/(r_H+x)^2+c_1+2c_2x]/(4\pi)\). A failure stops the shared ensemble; there is no separate histogram cut for each observable. The polynomial extends through the reference root, including negative \(x\). No interior state information is inferred from that extension.

At first order, the general covariance relations are
\begin{align}
\sigma_r^2&=C_{00}/a^2,\\
\sigma_T^2&=[C_{11}-2(b/a)C_{01}+(b/a)^2C_{00}]/(16\pi^2),\\
\operatorname{Cov}(\delta r_h,\delta T)&=-[C_{01}-(b/a)C_{00}]/(4\pi a).
\end{align}
Before finite-window projection, the differentiable model \(C(r,r')=s(r)s(r')e^{-(r-r')^2/(2L_c^2)}\) gives \(C_{00}=s^2\), \(C_{01}=ss'\), and \(C_{11}=s'^2+s^2/L_c^2\). Hence
\begin{equation}
\sigma_T^2=\frac{(s'-(b/a)s)^2+s^2/L_c^2}{16\pi^2}.
\end{equation}
This formula explicitly shows how envelope smoothing can change a temperature width without a comparable change of the anchored metric RMS. Numerical comparisons must use the projected coefficient covariance in Eq.~\eqref{eq:linear_matrix}, not silently replace its finite-window derivative by a point derivative.

\subsection{Reported statistics and their interpretation}
The histogram in main Fig.~3 has \(\langle r_h/r_H\rangle=0.9996\), \(\sigma_r/r_H=0.0185\), \(\langle T/T_H\rangle=1.0021\) and \(\sigma_T/T_H=0.0483\). Extended Data Fig.~1 uses the same 2,200 realizations: radius widths 0.0184 (linear) and 0.0185 (root-evaluated), and temperature widths 0.0482 and 0.0483. All values here are quoted at the precision displayed in the artwork.

For independent normal observations, \((N_s-1)\hat\sigma^2/\sigma^2\) is chi-square distributed with \(N_s-1\) degrees of freedom. Thus \(\operatorname{Var}(\hat\sigma^2)=2\sigma^4/(N_s-1)\), and the delta method gives \(\operatorname{SE}(\hat\sigma)/\sigma\simeq[2(N_s-1)]^{-1/2}\). At 2,200 samples this is 0.0151. The transformed root and temperature variables are only approximately Gaussian and the ensemble is centered as described above; this estimate is a reference scale rather than a measured confidence interval. A paired bootstrap of the shared coefficient rows would test the linear/root width difference while preserving its covariance, but no such bootstrap result is claimed in the supplied figures.

The root is a local candidate trapping horizon in the chosen spherical ansatz \cite{Hayward1994,AshtekarKrishnan2004,NielsenVisser2006}. It is not a globally determined event horizon. For the metric used above, \(\kappa_H=\tfrac12\Box_{(2)}r=\tfrac12(F'+F\psi')\), reducing to \(F'/2\) at the horizon \cite{Kodama1980,Hayward1998}. An asymptotically normalized stationary Killing surface gravity also involves \(e^{\psi_h}\); the present reconstruction sets \(\psi=0\). Predicting a detector spectrum or radiation-temperature distribution requires field propagation in an evolving geometry and cannot be inferred from the slope histogram alone.

\section{Resolution, exact identities and reproducibility}
The mode expansion of the exterior Unruh state, Wick's theorem for the free field, the stated angular averaging identities and covariance propagation by a specified linear map are exact mathematical relations. Their evaluation in the figures uses finite frequency and angular support, numerical mode extraction and quadrature. The density-driven gravitational closure, Gaussian radial correlation model, imposed amplitude and polynomial continuation are further assumptions. An exact identity at one stage does not make the combined numerical reconstruction an exact quantum-gravity solution.

For a smooth sampling function \(g(x)\), a smeared stress observable is \(\hat\rho[g]=\int\dd^4x\sqrt{-g}\,g(x)\hat\rho(x)\), with connected variance \(\int g(x)N_\rho(x,x')g(x')\,\dd V_x\dd V_{x'}\). This formulation identifies the resolution of a measurement and the distributional meaning of the noise kernel. The sharp frequency band used here is a stated numerical regulator, not proof that the unsmeared coincident variance is finite. The stress observable is quadratic: the field band \([\omega_{\min},\omega_{\max}]\) allows sum frequencies as high as \(2\omega_{\max}\) and difference frequencies down to zero. Cutting the stress integral at \(\Omega_{\max}=1.2/M\) therefore imposes an additional restriction even after the field band is fixed.

Numerical consistency is checked at several levels. A mode requires stable asymptotic coefficients as well as small ODE error. The bilocal stress spectrum must have the correct Hermiticity and nonnegative quadratic forms, not only nonnegative diagonal entries. The gravitational response requires a declared source component, slicing and outer condition. The reconstruction requires its actual covariance, fit window and root rule. Sampling refinement then addresses only the remaining finite-ensemble uncertainty. The agreement of main Fig.~3 and Extended Data Fig.~1 establishes consistency of the shared reconstruction; each source and response test addresses a separate numerical uncertainty.

The data and code supporting this study will be provided by the corresponding author upon reasonable request. A reproducible numerical record comprises the unnormalized response covariance, model covariance, coefficient samples, paired linear/root observables, seed, run parameters and mode diagnostics. These quantities permit a direct check of Eq.~\eqref{eq:linear_matrix} and a source-refinement study at fixed observable definition.